\documentclass[11pt,a4paper,bibliography=totoc,bibliography=openstyle, reqno]{amsart}
\usepackage{amsfonts}
\usepackage{amsthm}
\usepackage{amsmath}
\usepackage{amscd}
\usepackage{amssymb}
\usepackage{subfigure}
\usepackage[T1]{fontenc}            
\usepackage[utf8]{inputenc}       
\usepackage[english]{babel}                  
\usepackage{csquotes}               
\usepackage{t1enc}
\usepackage[mathscr]{eucal}
\usepackage{indentfirst}
\usepackage{graphicx}
\usepackage{graphics}
\usepackage{pict2e}
\usepackage{color}
\usepackage{epic}
\usepackage{bbm}
\usepackage{enumitem}
\usepackage{tikz}
\usepackage{tkz-graph}
\numberwithin{equation}{section}
\usepackage[margin=2.6cm]{geometry}
\usepackage{epstopdf} 
\usepackage{hyperref}
\usepackage{textcomp} 
\usepackage{autobreak}
\usepackage{subfigure}
\usepackage{float}
\usepackage{mwe}
\usepackage{bbold}
\usepackage{dsfont}
\usepackage{booktabs}
\usepackage{makecell}
\usepackage{diagbox}

\usepackage[backend=bibtex,               
          natbib=true,                 
          block = space,
           bibstyle=authoryear,        
           citestyle=authoryear,        
           babel=other,                 
           maxcitenames=2,              
           maxbibnames=10,              
           url=true,                   
           doi=true,                   
           giveninits=true]{biblatex} 
           
\renewbibmacro{in:}{} 
\DeclareNameAlias{sortname}{last-first} 
\DefineBibliographyStrings{german}{andothers={\mbox{et~al.}}} 
\newcommand{\Z}{\mathbb{Z}} 
\newcommand{\N}{\mathbb{N}}

\newtheorem{Theorem}{Theorem}[section]

\newtheorem{Prop}[Theorem]{Proposition}
\newtheorem{Algorithm}[Theorem]{Algorithm}

\newtheorem{Remark}[Theorem]{Remark}
\newtheorem{asum}{Assumption}
\newtheorem{?}[Theorem]{Problem}
\newtheorem{Example}[Theorem]{Example}

\newcommand{\psxn}{P_{S,x_n}}
\newcommand{\psxnp}{P_{S,x_n}^{(para)}}

\newcommand{\hpsxn}{\widehat{P}_{S,x_n}}

\newcommand{\hpsxnnp}{\widehat{P}_{S,x_n}^{(npara)}}

\numberwithin{equation}{section}

\begin{document}

\begin{center}
\begin{LARGE}
Predictive inference for discrete-valued time series
\end{LARGE}\\
\vspace{1cm}
\begin{Large}
Maxime Faymonville\footnote{Department of Statistics,  TU Dortmund University, D-44221 Dortmund, Germany; faymonville@statistik.tu-dortmund.de; corresponding author} \quad Carsten Jentsch\footnote{Department of Statistics,  TU Dortmund University, D-44221 Dortmund, Germany; jentsch@statistik.tu-dortmund.de} \quad Efstathios Paparoditis\footnote{Cyprus Academy of Sciences, Letters,  and Arts, CY-1011 Nikosia, Cyprus; stathisp@ucy.ac.cy}
\end{Large}
\end{center}

\vspace{1.5cm}

\begin{center}
    \textbf{Abstract}
\end{center}

For discrete-valued time series, predictive inference cannot be implemented through the construction of prediction intervals to some predetermined coverage level, as this is the case for real-valued time series. To address this problem, we propose to reverse the construction principle by considering preselected sets of interest and estimating the 
probability that a future observation of the process falls into these sets. The accuracy of the  prediction 
is then evaluated by quantifying the uncertainty associated with  estimation of these predictive probabilities. We consider parametric and non-parametric approaches and derive asymptotic theory for the estimators involved. 
Suitable bootstrap approaches to evaluate the distribution of the estimators considered  also are introduced.
They have the advantage to imitate the distributions of interest under different possible settings, including the practical important case where uncertainty holds true about the correctness of a parametric model used for prediction. 
Theoretical justification of the bootstrap is given, which also requires investigation of  asymptotic properties of  parameter estimators  under model misspecification.
We elaborate on bootstrap implementations under different scenarios and focus on parametric prediction using INAR and INARCH models and  (conditional) maximum likelihood estimators. 
Simulations investigate the finite sample performance of the predictive method developed and applications to real life data sets are presented. \\





\newpage

\section{Introduction}\label{sec:intro}
The class of discrete-valued time series models is vast;  see \citet{bookweiss} and \citet{davis_handbook} for an overview. A first family  of such time series  are so-called count time series which include, e.g., the INAR and the INARCH models that will be further  discussed in Section \ref{sec:inar1}. A second family are categorical time series (which also include ordinal time series), when the data exhibit a qualitative range consisting
of a finite number of categories; see e.g., \citet{pruscha}, \citet{fok_categ},  \citet{weiss_ord} and \citet{ordinal_airq}. A special case of categorical time series are binary time series, where  the observations take two possible values only;  see e.g. \citet{kedem_binary} and \citet{lena}. 

In the continuous  time series setting, predictive inference commonly deals  with  point prediction which is    accompanied by a  prediction interval, where the goal of the latter  is to  take into  account  the uncertainty associated with the point prediction. Given time series data up to some time point $n$, that is, given  $ X_1, \ldots, X_n$, such prediction intervals are  designed to cover future observations, say  $X_{n+h}$, for some $ h \in \N$,  with a desired (high)  probability. The main   criterion to judge the quality of such  prediction intervals is  its asymptotic validity, which is specified, e.g., in \citet{pandp} as follows. Given  time series  $X_1, \ldots, X_n$, suppose that the  task is to predict one step ahead, that is,  to predict $X_{n+1}$.  Then, for $\beta \in(0,1)$, the interval $[l_n, u_n]$ is called an asymptotically valid $(1-\beta)$ prediction interval for $X_{n+1}$ if
\begin{align} \label{eq:asy_val}
 P(l_n \leq X_{n+1} \leq u_n |X_1, \ldots, X_n) \rightarrow 1-\beta \quad \text{for} \;  n \rightarrow \infty.
\end{align} 
In the formulation above,  $l_n=l_n(X_1,\ldots,X_n)$ and $u_n=u_n(X_1,\ldots,X_n)$  typically are functions of the observed sample $X_1, \ldots, X_n$.

In the context  of discrete-valued time series, the above  concept of an (asymptotically) valid prediction interval is not applicable since  \emph{intervals} do not account for the discrete nature of the data. In fact and for general choices of $\beta \in (0,1)$, it  is typically not possible to construct an interval which guarantees an exact (or asymptotic)  level $1-\beta$.  To address this problem, in the special case of count time series, \citet{coh_forecast} introduced the notion  of \emph{coherent forecasting} according to which predicted values of count processes should also be counts themselves. In compliance with this notion, we could  rather consider prediction \emph{sets} than intervals. These can be  obtained by looking at the set  $[l_n,u_n]\cap range(X_{n+1})$. However,  also for such  prediction sets, it is in  general  not even asymptotically possible to achieve validity in the sense of \eqref{eq:asy_val}. This is  illustrated by the following  example.

\begin{Example}
  Let $X_1, \ldots, X_n$  be an ordinal stationary time series stemming from a first-order Markov process with $range(X_t)=\{0,1, \ldots, 5\} $.  Think of $ X_t$  as measuring  the daily   air quality level in different cities; see  the data example of  \citet{ordinal_airq} and the analysis  by \citet{jahn_weiss}. Suppose that for some $x_n \in \{0,1,\ldots, 5\}$, e.g., for $x_n=2$,  the process has the following one step transition probabilities: $P(X_{n+1}=0 | X_n = 2)=0.58$, $P(X_{n+1}=1| X_n = 2)=0.2$, $P(X_{n+1}=2| X_n = 2)=0.11$, $P(X_{n+1}=3| X_n = 2)=0.09$ and  $P(X_{n+1}=4| X_n = 2)=P(X_{n+1}=5| X_n = 2)=0.01$. Then, it is not possible to obtain a prediction set for $X_{n+1}$ given that $ X_n=2$ with an exact or asymptotic coverage of $95\%$; see Table \ref{tab:example} for some exemplary prediction sets.
\end{Example} 

\begin{table}[h]
\centering
\begin{tabular}{c|cc|cc|cc|cc}
$X_{n+1}\in$ & Set  & Cov & Set  & Cov & Set   & Cov & Set  & Cov  \\ 
  \hline
 & $\{1,2,3\}$ & 0.40  & $\{0,1,2\}$ & 0.89 
   & $\{0,1,2,4,5\}$ & 0.91 
  & $\{0,1,2,3\}$ & 0.98 
\end{tabular}
\vspace{0.5cm}
\caption{Coverages (Cov) for different exemplary prediction sets.}
\label{tab:example}
\end{table}

A way to achieve coherent forecasting for the setup of count processes  is to return the full predictive probability mass function of $ X_{n+h}$ given $X_1, \ldots, X_n$, as  proposed by \citet{homb_pmf}. However, this does not account for the estimation uncertainty that comes with it. 
In this paper, we propose an alternative strategy to solve the problem of coherent forecasting by transforming the predictive inference problem into a parameter estimation problem and by  constructing  a corresponding \emph{confidence} interval for the underlying parameter  at some  desired  level $1-\delta$, where  $\delta\in(0,1)$. 
To be more specific, let $ S \subset \mathbb{R}$ be any user-selected subset of possible values of interest that $X_{n+h}$ can take. We propose to estimate the predictive probability of this set, that is, the probability that $X_{n+h}$ takes a value in $S$ given the observed stretch $X_1,\ldots, X_n$ of the process. The uncertainty associated with such a  prediction can then be accounted for by constructing  a $(1-\delta)$-confidence interval for the corresponding   predictive probability. Clearly,   if  $range(X_{n+h})$ is known,   $ S\subset range(X_{n+h})$ is  a typical choice, which we assume to be the case for simplicity in what follows. Note that  $ |S|=\infty$ is also allowed. 

In Section \ref{sec:main}, we elaborate on this  general idea. In Section~\ref{sec:asym}, we   derive  some asymptotic theory for the corresponding prediction problem, where we concentrate  on   parametric and  non-parametric implementations  of this kind of  predictive inference. As we will see,  using the asymptotic distributions derived for the  calculation of  the described  confidence intervals turns out to  be    cumbersome in practice. To circumvent this problem, we  resort to  bootstrap  techniques. 
In particular, 
we  use bootstrapping to construct \emph{confidence} intervals of the predictive  probabilities of interest under a variety of relevant settings. In   Section \ref{sec:bootstrap}, we propose  different bootstrap algorithms  and  we also address the important problem of model uncertainty (in case the decision of the user is in favor of a parametric model)  by providing  a bootstrap procedure which takes this problem into account in constructing the confidence interval of interest.  To derive theoretical results  in this context,  the  investigation of the  consistency and the distributional properties of  estimators under model misspecification is required, a topic which is of interest on its own.  In Section \ref{sec:inar1}, we concentrate on  asymptotic and on bootstrap-based approaches for the case of (discrete-valued) count time series, where predictive inference is implemented using   (conditional) maximum likelihood estimation and (first-order) INAR and INARCH models.  Section \ref{sec:pie} discusses some practical issues while  Section \ref{sec:sims} presents   simulation results demonstrating the capabilities of the different asymptotic and bootstrap methods proposed.  Section \ref{sec:appl} gives applications to  real-life  data sets  and   Section \ref{sec:concl} concludes the findings of  our paper. All technical proofs are deferred to the Appendix.  

\bigskip

\section{Prediction for Discrete-Valued Time Series} \label{sec:main}

\subsection{Predictive Probabilities} Suppose we observe a sample $X_1, \ldots, X_n$ from  a strictly  stationary, discrete-valued process $(X_t,t\in\Z)$, that is  $ range(X_0) $  $ \subset {\mathbb R}$ is a countable set. Without loss of generality, let  $range(X_0)=\{y_k,k\in\mathcal{N}\}$, where 
$\mathcal{N} \subseteq \mathbb{N}$.
Our goal is to predict the future value  $ X_{n+h}$  of the process for some $ h \in \N$.  In this  setup and  for any  $ S \subset \mathbb{R}$, we denote  by 
$$ P_{S,\underline{x}_n}^{(h)}= P(X_{n+h} \in S \big| \underline{X}_n=\underline{x}_n) $$ 
the {\it $h$-step predictive probability} of the subset $S $ given that \[ \underline{X}_n:=(X_1, \ldots, X_n) = (x_1,  \ldots, x_n)=:\underline
{x}_n. \]
$P_{S,\underline{x}_n}^{(h)} $ is the conditional probability that the future random variable $ X_{n+h} $ takes a value in the subset $S$ given the realization  $\underline{x}_n$ of $\underline{X}_n$. Let $ \delta \in (0,1)$ be small. A $(1-\delta)$-confidence interval for $ P_{S,\underline{x}_n}^{(h)}$, that is, an interval $ [L_n^{(h)}, U_n^{(h)}]$  with the property 
$$ P\big(L_n^{(h)} \leq P_{S,\underline{x}_{n}}^{(h)} \leq U_n^{(h)} \big) =1-\delta,$$
 is called a {\it $(1-\delta)$ confidence interval for the $h$-step predictive probability} of   the subset $S$. If the above statement holds true for  $n\rightarrow \infty$, that is, if $$\lim_{n\rightarrow\infty}P\big(L_n^{(h)} \leq P_{S,\underline{x}_{n}}^{(h)} \leq U_n^{(h)} \big) =1-\delta,$$ the corresponding interval is called an {\it asymptotic $(1-\delta)$ confidence interval for the $h$-step predictive probability} of the subset $S$.


In the following, we concentrate on the case where the underlying process is a first-order, discrete-valued  Markov process and   $h=1$.  In this case  $P_{S,\underline{x}_n}^{(h)} =P_{S,x_n}^{(h)} $ and we  write for simplicity, since  $h=1$,
\begin{align*}
\psxn := P(X_{n+1} \in S | X_n = x_n) = \sum_{k\in \mathcal{N}: y_k\in S} P(X_{n+1}=y_k | X_n=x_n). 
\end{align*}
We assume throughout this paper that $ x_n \in range(X_0)$  is a given value which coincides with  the value of the  last observation of the time series $X_1,  \ldots, X_n$ at hand.  Furthermore and in order to simplify notation, we sometimes also write $ P_{X_{t+1}=y_k|X_t=x_n}$ for the one step transition probability $ P(X_{t+1}=y_k|X_t=x_n)$. Note that our derivations  can  easily be extended to Markov models of higher order and to larger prediction horizons, that is,  $h>1$; we refer to Section~\ref{subsec:extension} for more details.  We start by imposing the following assumption.

\begin{asum}[Markov process] \label{assum_np} 
 $(X_t, t \in \mathbb{Z})$ is a discrete-valued, homogeneous, aperiodic, irreducible, positive recurrent and geometrically ergodic  first-order Markov process.
\end{asum}

\begin{Remark}
Assumption \ref{assum_np} covers a wide range of discrete time series. It even includes appropriate nominal time series since we do not require an order of the potentially involved categories of the data. An example for the latter time series are DAR processes; see \citet{dar_model} for details.
\end{Remark}

As already mentioned, our  aim is to construct a confidence interval for the (one step ahead) predictive probability of   a  set $S\subset \mathbb{R}$  of interest, which has an (asymptotic)  coverage of  $1-\delta$ for  some given $ \delta \in (0,1)$.  
In  the next two subsections, we will consider parametric and non-parametric approaches towards this goal.

\subsection{Asymptotic Confidence Intervals} \label{sec:asym}
\subsubsection{Parametric Prediction} \label{subsec:main_p}
A common  situation in time series analysis  occurs when a parametric model  is used to perform  prediction. 
In our context,  the predictive probability $ \psxn $ is  estimated using   a parametric,  first-order Markov model, the properties of which will be specified    in   Assumption~\ref{assum_p} below. In order to emphasize the dependence of the prediction on the used model  and on  the associated parameter vector denoted by $\theta$, we write  $ (X_t(\theta),t\in\Z)$ for the parametric model used for prediction. 
Note  that  we  do not want to  assume that the discrete-valued  time series observed  necessarily stems from the parametric process   $(X_t(\theta),t\in\Z)$ which will be  used for prediction. This seems important in order  to   address the  practically important situation  of  model misspecification in implementing the prediction  and in constructing the  confidence   interval for the  predictive probability. This situation  occurs  when the parametric model class used for prediction does not necessarily coincide with  the data generating process.  The following (high-level)  assumption   specifies our   requirements  including those needed to obtain an asymptotically valid confidence interval for the predictive probability of interest. 

\begin{asum}[Parametric family of Markov processes] \label{assum_p} \phantom \\
\begin{itemize}
\item[(i)] ${\mathcal M}_\theta:= \{(X_t(\theta), t \in \mathbb{Z}), \theta\in\Theta\}$ is a parametric family  of stationary, discrete-valued, aperiodic, positive recurrent  and geometrically ergodic, first-order  Markov process with parameter vector  $\theta \in \Theta \subset  \mathbb{R}^d$ for some finite $d \in \mathbb{N}$ and  $ \Theta $  a compact set. Furthermore, $\theta_1\neq \theta_2$ implies   $ (X_t(\theta_1),t\in\Z) \neq (X_t(\theta_2),t\in\Z)$.
\item[(ii)] There exists a unique $\theta_0\in\Theta$ such that, when fitting a model from the family ${\mathcal M}_\theta$ to the process $(X_t,t\in\Z)$, the estimator $\widehat \theta$ used fulfills
\begin{align*}
\sqrt{n}(\widehat \theta - \theta_0) \overset{d}{\rightarrow} \mathcal{N}(0, V_{\theta_0}),
\end{align*}
where $V_{\theta_0}$ is a positive-definite  covariance matrix.
\end{itemize}
\end{asum}

Under the above  assumption, we write   $ \psxn^{(para)}(\theta_0):=\psxn$ for the (parametric) predictive probability of the set $S$ when model class $\mathcal{M}_\theta$ 
is used for prediction.  Then, 
$$ P^{(para)}_{S,x_n}(\widehat{\theta})=\sum_{y_k\in S} P^{(para)}_{X_{t+1}=y_k|X_t=x_n}(\widehat{\theta}),$$ is  the estimator of  $ P_{S,x_n}^{(para)}(\theta_0)$ obtained by replacing $ \theta_0$ by $ \widehat{\theta}$. Here, $P^{(para)}_{X_{t+1}=y_k|X_t=x_n}(\theta_0) $ denotes the one step transition probability associated with the parametric Markov process $ (X_t(\theta_0), t\in\Z) \in {\mathcal M}_\theta$.
Now assuming that $P^{(para)}_{X_{t+1}\in S|X_t=x_n}(\theta) $ is continuous differentiable with respect to $\theta$, we get 
\[ \sqrt{n}\big(P^{(para)}_{X_{t+1}\in S|X_t=x_n}(\widehat{\theta}) -P^{(para)}_{X_{t+1}\in S|X_t=x_n}(\theta_0) \big)= \nabla_\theta P^{(para)}_{X_{t+1}\in S|X_t=x_n}(\theta^\prime)\sqrt{n}(\widehat{\theta}-\theta_0)^\top \]
for some $ \theta^\prime$ such that $ \|\theta^\prime-\theta_0\|\leq \|\widehat{\theta}-\theta_0\|$. Here and for any $ \theta_1\in \Theta$, $ \nabla_\theta P^{(para)}_{X_{t+1}\in S|X_t=x_n}(\theta_1)$  denotes the $d$-dimensional vector of partial derivatives of $P^{(para)}_{X_{t+1}\in S|X_t=x_n}(\theta) $ with respect to $\theta$ and evaluated at $ \theta_1$. We then get by Assumption \ref{assum_p}  and assuming  continuity of the partial derivatives, that as $n\rightarrow \infty$, 
\begin{align} \label{eq.ConvNor}
\sqrt{n}\big(P^{(para)}_{X_{t+1}\in S|X_t=x_n}(\widehat{\theta})  -P^{(para)}_{X_{t+1}\in S|X_t=x_n}(\theta_0) \big)
\overset{d}{\rightarrow} \mathcal{N}\Big(0, \sigma^2_S(\theta_0)\Big),
\end{align}
where 
\[\sigma_S^2(\theta_0):=\nabla_\theta^\top  P^{(para)}_{X_{t+1}\in S|X_{t}=x_n} (\theta_0)\,V_{\theta_0}\,\nabla_\theta P^{(para)}_{X_{t+1}\in S|X_t=x_n}(\theta_0).\] 

The following result  easily follows. 

\begin{Prop}[Asymptotic parametric confidence interval for $P^{(para)}_{S, x_n}(\theta_0)$] \label{prop22}
    Suppose Assumption \ref{assum_p} holds true and let $V_{\widehat \theta}$ be a consistent estimator of  $V_{\theta_0}$ obtained by replacing $ \theta_0$ by $\widehat{\theta}$. Suppose that  $P^{(para)}_{S,x_n}(\theta)$ is continuously differentiable around  $\theta_0$ and that  $\nabla_\theta \psxn^{(para)}(\theta_0) \neq 0$. Then, 
    \[ \left[\psxn^{(para)}(\widehat \theta) - \frac{z_{\delta /2} \sigma_S(\widehat \theta)}{\sqrt{n}} , \psxn^{(para)}(\widehat \theta) +  \frac{z_{\delta /2} \sigma_S(\widehat \theta)}{\sqrt{n}} \right] \]
    is an asymptotically valid confidence interval for the predictive probability $ P^{(para)}_{S, x_n}( \theta_0) $ with confidence level $1-\delta$, where  $\sigma_S(\widehat \theta) =\sqrt{\sigma^2_S(\widehat \theta)} $ and $\sigma_S^2(\widehat{\theta})$
    is the estimator of $\sigma_{S}^2(\theta_0)$
obtained by replacing $\theta_0$ by $\widehat{\theta}$.     
  Furthermore,   
   $ z_{\delta/2}$ denotes the upper $ \delta/2$-quantile of the standard Gaussian distribution.
\end{Prop}

\begin{Remark}{~}
\begin{itemize}
\item[(i)]  Note that in the case of model misspecification, which is allowed   by Assumption~\ref{assum_p},  $\theta_0$ is the parameter vector from the parameter space $\Theta$ which best fits the underlying process $(X_t,t\in\Z)$ and that the particular value of $ \theta_0$  also depends on the  estimation method used to obtain 
$\widehat{\theta}$.  Furthermore, the same will be true 
for the (limiting) distribution of the estimator $ \widehat{\theta}$ and in particular for the variance $ V_{\theta_0} $ of this limiting distribution. 
We elaborate in  Section \ref{sec:inar1} on   the important case where $ \widehat{\theta}$ is a (conditional) maximum likelihood estimator. 
\item[(ii)]  The construction of the asymptotic  confidence interval for the predictive probability $ \psxn^{(para)}(\theta_0)$ relies on  the estimator $\sigma^2_S(\widehat \theta)$, which is typically  difficult to obtain  in practice due to the need to calculate the unknown quantities   $\nabla_\theta \psxn^{(para)}(\widehat\theta)$ and $ V_{\widehat{\theta}} $. In Section \ref{sec:bootstrap}, we will introduce  bootstrap procedures to estimate these quantities as well as the distribution of $P_{S,x_n}^{(para)}(\widehat{\theta})$. 
\item[(iii)]
 Assumption \ref{assum_p} (ii)  can be generalized to allow   for an infinite dimensional parameter space $\Theta$. For example, this is relevant for the semi-parametric INAR model investigated by \citet{drost} and \citet{faymonville2025}. 
 \end{itemize}
\end{Remark}

\subsubsection{Non-parametric Prediction} \label{subsec:main_np}
Alternatively  to the situation discussed in Section~\ref{subsec:main_p}, we may consider a fully non-parametric approach  to estimate the predictive probability  and to obtain  the corresponding confidence interval. In the absence of  parametric assumptions, we  can  use the appropriate relative frequencies in order to estimate  $ \psxn$. In the following, we denote the predictive probability $ \psxn$ by $\psxn^{(npara)} $ in order to distinguish it from the parametric case discussed in the previous section. A  natural estimator of $\psxn^{(npara)} $
is then obtained as follows:
\begin{align} \label{eq.EstofConPro}
\widehat{P}^{(npara)}_{X_{t+1}\in S|X_t=x_n} = \left\{ \begin{array}{lll} \frac{\sum \limits_{t=1}^{n-1} \mathbf{1}_{\{ X_{t+1} \in S, X_t = x_n \}}}{\sum \limits_{t=1}^{n-1} \mathbf{1}_{\{ X_t = x_n \}}} & & \mbox{if} \ \sum_{t=1}^{n-1} \mathbf{1}_{\{ X_t = x_n \}} \neq 0, \\
& & \\
0 & & \mbox{if}\  \sum_{t=1}^{n-1} \mathbf{1}_{\{ X_t = x_n \}} =0.
\end{array} \right.
\end{align}
Notice that by Assumption \ref{assum_np},   the estimator  $\widehat{P}^{(npara)}_{X_{t+1}\in S|X_{t}=x_n}$
   converges in probability to $ P_{X_{t+1}\in S|X_{t}=x_n}$ as $n\rightarrow\infty$
  ; see \citet{derman1956}, where also a limiting Gaussian distribution for these estimators has been established. 
An asymptotically valid $(1-\delta)$-confidence interval for $\psxn^{(npara)}$ is  given in   the   following proposition.

\begin{Prop}[Asymptotic non-parametric confidence interval for $P_{S, x_n}=\psxn^{(npara)}$] \label{theo:ci_np}
    Suppose that  Assumption \ref{assum_np} holds true. Then,  an asymptotically valid confidence interval  for $\psxn^{(npara)}$ with confidence level $1-\delta$ is given by
\[  \left[\hpsxnnp - \frac{z_{\delta/2} \widehat\sigma_S}{\sqrt{n-1}},   \hpsxnnp + \frac{z_{\delta/2} \widehat\sigma_S}{\sqrt{n-1}}\right],  \]
where $\widehat\sigma_S =\sqrt{\widehat\sigma^2_S} $ with $\widehat \sigma_S^2=\widehat a_S^T \widehat \Sigma_S \widehat a_S$. Here, using the notation $\widehat Q_{S,x_n}=\frac{1}{n-1}\sum_{t=1}^{n-1}\mathbf{1}_{ \{X_{t+1} \in S, X_t=x_n \}}$ and $\widehat Q_{x_n}=\frac{1}{n-1}\sum_{t=1}^{n-1}\mathbf{1}_{ \{ X_t=x_n \}}$, we have
\begin{align*}
    \widehat a_S = \Bigg(\frac{1}{\widehat Q_{x_n}},\frac{-\widehat Q_{S,x_n}}{\widehat Q_{x_n}^2}\Bigg)^\top  \quad   \text{and}  \quad     \widehat \Sigma_S = \begin{pmatrix}
\widehat \Sigma_S^{(1,1)}  & \widehat \Sigma_S^{(1,2)}   \\
\widehat \Sigma_S^{(2,1)} & \widehat \Sigma_S^{(2,2)}
\end{pmatrix},
\end{align*}
where $\widehat \Sigma_S$ is a consistent estimator for $\Sigma_S$ defined in \eqref{eq:Sigma}.
The  estimator $\widehat \Sigma_S$  is obtained by replacing the corresponding probabilities in $\Sigma_S$ by their sample estimators and suitably truncating the corresponding infinite sums.

\end{Prop}


\subsection{Bootstrap  Confidence Intervals} \label{sec:bootstrap}

Despite the fact whether a parametric or a non-parametric approach is used to estimate  the predictive probability, when it comes to the construction of the corresponding  confidence interval,   some care is needed. To elaborate, recall that    parametric assumptions that have been imposed  on  the underlying process in order to obtain estimators of the predictive probabilities, may  not  necessarily hold true in reality. Depending on which  constellation  holds true,  this (may) affect the  (limiting) distribution of the estimators  used and therefore, it has to   be  taken into account  in order to properly construct the  confidence interval of interest.  In Table~\ref{tab:T1}, we  show the different  constellations that may occur when it comes to prediction.  This  table   summarizes  different situations    depending on whether  the time series  $X_1, \ldots, X_n$ observed stems from a  particular parametric class of  Markov processes denoted by  $ {\mathcal M}_\theta$ or not and whether  a parametric or a non-parametric approach is used to perform  prediction, that is, to  estimate  the predictive probability $ P_{S,x_n}$.

\vspace*{0.3cm}

\begin{table} 
\begin{tabular}{c|c|c}
 
 \diagbox[]{Prediction}{Data} & \makecell{$X_1, \ldots, X_n$ \\   
    stems from the  class  $ {\mathcal M}_\theta$} &  \makecell{ $X_1, \ldots, X_n$ \\ doesn't stem from the   class $ {\mathcal M}_\theta$} \\ 
\hline 
\makecell{Non-parametric  \\
 $\widehat{P}^{(npara)}_{S, x_n}$}  & \makecell{Model is ignored \\ } & \enquote{Correct} prediction \\
\hline 
\makecell{Parametric  \\
$P^{(para)}_{S, x_n}(\widehat{\theta})$}  & \enquote{Correct} prediction & Model is misspecified \\
\end{tabular}
\vspace*{0.5cm}
\caption{Possible constellations that may occur when using a parametric or non-parametric estimator of predictive probabilities.}
\label{tab:T1}
\end{table}

As it can be seen from this table, there are two constellations where the  particular method used to calculate the predictive probability is correct and  which  are termed as    "Correct" Prediction. The first    refers to the situation where the time series stems from the parametric class $ {\mathcal M}_\theta$ and the same  class  also is used to calculate the predictive probability. The second concerns the case where the time series does not stem from the  parametric class $ {\mathcal M}_\theta$ and the prediction is implement in a non-parametric way. In the other two constellations, either an error is made in performing the prediction or the approach is inefficient. To elaborate, an error occurs in the case where the parametric model class $ {\mathcal M}_\theta$ is wrongly used (model misspecification). Analogously,    the predictive approach becomes inefficient when the parametric model class $ {\mathcal M}_\theta$ is erroneously  not used and  the prediction is implemented in a non-parametric way. Now, in order to  construct a proper confidence interval for the predictive probability of interest, these different constellations have to be taken into account. 

The bootstrap procedures  developed in this section   for the  construction of   confidence intervals   are able to fully take into account the different situations that may occur as described in Table~\ref{tab:T1}. Additionally to this  important fact, the bootstrap also has  the  advantage to circumvent the possibly cumbersome calculations needed to implement in practice the limiting distributions derived  in  Sections \ref{subsec:main_p} and \ref{subsec:main_np} in order to  construct the confidence intervals of interest.

The main  structure  of the  bootstrap procedure proposed consists of four  steps, where  the first two have to be  specified differently  according to which setup from the four possible ones shown in Table~\ref{tab:T1}  should be  imitated in the bootstrap world.  We  first state  this basic bootstrap algorithm, while   the  different specifications of the first two steps that  should be  made will be discussed later on in more detail. 
 Notice that in the following description, $\widehat{P}_{S,x_n}$, respectively, $ \widehat{P}_{S,x_n}^\ast$, are used to denote any estimator, respectively,  bootstrap estimator,  of the predictive probability $ P_{S,x_n}$, where   the specific form these estimators  take    will be clarified in the discussion  following the basic bootstrap algorithm.

\begin{Algorithm}[Basic bootstrap algorithm for the construction of  confidence intervals  of predictive probabilities] \label{algo_bs} \phantom \newline
\begin{itemize}

\item[Step 1:] Given $X_1,\ldots, X_n$, calculate the estimator $ \widehat{P}_{S,x_n}:=\widehat{P}_{S,x_n}(X_1,\ldots,X_n)$ of the predictive probability $ P_{S,x_n}$.

\item[Step 2:] Generate  a pseudo time series $ X_1^\ast, \ldots, X_n^\ast$ and calculate the same estimator as in Step 1 but based on $X_1^\ast,\ldots,X_n^\ast$ to get $ \widehat{P}_{S,x_n}^\ast:=\widehat{P}_{S,x_n}(X_1^\ast,\ldots,X_n^\ast)$.

\item[Step 3:] Repeat Step 2 a large number of times, say  $B$ times,  and calculate 
\begin{align*}
L_n^{*,(b)} = \widehat{P}^{\ast,(b)}_{S,x_n} - \widehat{P}_{S,x_n}, \; b=1, \ldots, B,
\end{align*}
where $ \widehat{P}^{\ast,(b)}_{S,x_n}$ denotes the estimator  $ \widehat{P}_{S,x_n}^\ast$ obtained in  the $b$th  bootstrap repetition.  
\item[Step 4:] Compute the $(1-\delta)$-confidence interval  as 
\begin{align*}
\left[  \hpsxn - q^*_{1-\delta/2} , \hpsxn - q^*_{\delta/2}   \right], 
\end{align*}
where $q_\alpha^*$ denotes the $\alpha$-quantile of the empirical distribution of $L_n^*$, that is, $ P^*(L_n^\ast \leq q^\ast_\alpha)=\alpha$. 
\end{itemize}
\end{Algorithm}

Recall  that we aim to imitate the distribution of the   estimator $\widehat{P}_{S,x_n} $ of the predictive probability  under the different possible scenarios. So far,  we have considered  two types of estimators: the  non-parametric estimator, that is, $ \widehat{P}_{S,x_n} = \widehat{P}_{S,x_n}^{(npara)}$ and the parametric estimator, that is,    
$ \widehat{P}_{S,x_n}=P^{(para)}_{S,x_n}(\widehat{\theta})$. Now, depending  on whether the time series $ X_1,\ldots, X_n$ stems from the parametric model class ${\mathcal M}_\theta$   or not, the distribution of the estimators $\widehat{P}_{S,x_n}^{(npara)} $, respectively, $P^{(para)}_{S,x_n}(\widehat{\theta}) $ may be different. 
This requires a proper specification of Step 1 and Step 2 of the basic bootstrap algorithm, which is clarified in   Table~\ref{tab:T2}.

\begin{small}
\begin{table}[t] 
\begin{tabular}{c|c|c}
 
 \diagbox[]{Prediction}{Data} & \makecell{$X_1, \ldots, X_n$ \\   
    stems from the class  $ {\mathcal M}_\theta$} &  \makecell{ $X_1, \ldots, X_n$ \\ doesn't stem from the class $ {\mathcal M}_\theta$} \\ 
\hline 
\makecell{\textit{Step 1}: \\ Non-parametric  \\
 $\widehat{P}^{(npara)}_{S, x_n}$}  & \makecell{ {\it Step 2:}   $ X_1^\ast, \ldots, X_n^\ast$  is   
 generated \\ \ \ \ \ using the estimated, model-based \\  \ \ \ \ one step transition probabilities \\  \ \ \ \ and   $ \widehat{P}^\ast_{S,x_n}= \widehat{P}_{S,x_n}^{\ast,(npara)} $ \ \ \ \ \ \ \ \ } 
& \makecell{ {\it Step 2:}   $ X_1^\ast, \ldots, X_n^\ast$  is   
 generated \\ using the non-parametrically \\
 estimated  one step transition \\  \hspace{1cm} probabilities 
 and   $ \widehat{P}^\ast_{S,x_n}= \widehat{P}_{S,x_n}^{\ast,(npara)} $\ \ \ \ \ \ \ \ } \\
 & & \\
\hline 
\makecell{\textit{Step 1}: \\ Parametric  \\
$P^{(para)}_{S, x_n}(\widehat{\theta})$}  & \makecell{ {\it Step 2:}   $ X_1^\ast, \ldots, X_n^\ast$  is   
 generated \\  using the estimated, model-based \\ one step transition probabilities \\ 
 and   $ \widehat{P}^\ast_{S,x_n}= P_{S,x_n}^{(para)}(\widehat{\theta}^\ast) $} 
 &  \makecell{ {\it Step 2:}   $ X_1^\ast, \ldots, X_n^\ast$  is   
 generated \\ using the non-parametrically \\
 estimated  one step transition \\  probabilities and    $ \widehat{P}^\ast_{S,x_n}= P_{S,x_n}^{(para)}(\widehat{\theta}^\ast)  $} 
\end{tabular}
\vspace*{0.5cm}
\caption{Different  possible specifications of Step 1 and Step 2 of the basic bootstrap algorithm depending on which  one of the four possible scenarios in Table~\ref{tab:T1} should   be imitated.}
\label{tab:T2}
\end{table}
\end{small}


\begin{Remark}[Unconditional confidence interval vs. conditional prediction interval] \label{rem-gen}
In contrast to bootstrap procedures   that aim  to  construct  (conditionally) valid prediction intervals  in the continuous case, we do explicitly not require that the pseudo time series $ X_1^\ast,  \ldots, X^\ast_n$ (in all four cases) has the property that $ X_n^\ast=x_n$ holds true. This is because in the discrete setting considered in this paper, we propose to reverse the approach and aim for the construction of a \emph{confidence} interval for the predictive probability $P_{S,x_n}$ instead of constructing a \emph{prediction} interval for $X_{n+1}$ (conditional on $X_n=x_n$).
\end{Remark}

In Sections \ref{subsec:bs_p} and \ref{subsec:bs_np}, we will discuss in more detail the implementation and the properties of the different specifications of the basic bootstrap algorithm described in   Table~\ref{tab:T2}.  Notice that  investigations of the asymptotic validity of the constructed bootstrap prediction intervals essentially need to show that the prediction error
 $L_n := \hpsxn - \psxn$ and the bootstrap prediction error $L_n^* := \hpsxn^{*} - \hpsxn$ converge, in a proper way, to the same  limiting distribution, where the latter is assumed to be continuous.  Denoting by  $q_\alpha$ the $\alpha$-quantile of the distribution of $L_n$, this convergence would justify the use of the bootstrap for constructing the prediction intervals of interest due to  the  following approximation
\begin{align*}
1- \delta &= P(q_{\delta/2} \leq L_n \leq q_{1-\delta/2})\\ 
& = P(\widehat{P}_{S,x_n}  - q_{1-\delta/2}  \leq P_{S, x_n}  \leq  \widehat{P}_{S,x_n}  - q_{\delta/2} ) \\
& \approx P(\widehat{P}_{S,x_n}  - q^*_{1-\delta/2}  \leq P_{S, x_n}  \leq  \widehat{P}_{S,x_n}  - q^*_{\delta/2} ).
\end{align*}
Here, $q_\alpha^*$ denotes  the $\alpha$-quantile of the  distribution of the bootstrap error  $L_n^*$, which, by the assumed convergence in distribution and the continuity of the limiting distribution,  satisfies    $ |q^\ast_{\delta/2} -  q_{\delta/2}| \rightarrow 0$, in probability, as $n\rightarrow\infty$. 

\subsubsection{Parametric Prediction} \label{subsec:bs_p}
In the case of  parametric prediction, that is, in the case  where the estimator $ P_{S,x_n}^{(para)}(\widehat{\theta})$  is used,   the following two different constellations  may occur as shown in the second row of Table~\ref{tab:T2}:
\begin{enumerate}
\item[(a)] \ $X_1,  \ldots, X_n$ stems from the parametric class used to perform the prediction.
\item[(b)] \ $ X_1,  \ldots, X_n$ does not stem from the parametric class used to perform the prediction.
\end{enumerate}

We  first elaborate on case (a).  Under this scenario, the matrix
of transition probabilities used to generate $ X_1^\ast,  \ldots, X_n^\ast$, is  obtained from  the parametric model and depends on the estimated parameter  vector $\widehat\theta $. Its $(i,j)$th
 element is given by $P^{(para)}_{X_{t+1}=x_j|X_t=x_i}(\widehat\theta)$.
The bootstrap pseudo time series $X_1^\ast,  \ldots, X_n^\ast$ is then generated  using these parametric one step transition probabilities. 
 
In case  (b),  the bootstrap time series $ X_1^\ast, \ldots, X_n^\ast$ is generated using non-parametric estimators  of the one step  
transition probabilities. The $(i,j)$th element of the corresponding transition probability matrix is $\widehat P^{(npara)}_{X_{t+1}=x_j|X_t=x_i}$ and it is given  in  \eqref{eq.EstofConPro} for  $S=\{x_j\}$ and $x_n=x_i$.

However, in any one of the two scenarios discussed,  the bootstrap estimator $ \widehat{\theta}^\ast$ based on  the pseudo time series $ X_1^\ast, \ldots, X_n^\ast$, has to imitate the distribution of the estimator $ \widehat{\theta}$ as stated in  the following  assumption. 
This high-level assumption has to be validated in a particular setting of interest, as this will be discussed  in more detail in Section \ref{sec:inar1}.

\begin{asum}[Bootstrap CLT for parameter estimator] \label{asum_p_bs}
    The bootstrap estimator $ \widehat{\theta}^\ast$ based on the pseudo time series $ X_1^\ast,  \ldots, X_n^\ast$ generated under the two different constellations (a) or (b), satisfies  
    \[ \sqrt{n}(\widehat{\theta}^* - \widehat\theta) \overset{d}{\rightarrow} \mathcal{N}(0, V_{\theta_0}) \; \text{in probability}, \]
    where $ V_{\theta_0}$ is given in Assumption 2(ii). 
\end{asum}

Under this assumption, we can  establish validity of the bootstrap procedure used to construct   confidence intervals for the predictive probability in case the parametric estimator $ P^{(para)}_{S,x_n}(\widehat{\theta})$ is used.

\begin{Theorem}[Parametric bootstrap confidence interval for $P^{(para)}_{S, x_n}(\theta_0)$]
    Suppose Assumptions \ref{assum_np}, \ref{assum_p} and \ref{asum_p_bs} hold true.
Let $\psxn^{(para)}(\theta)$ be continuously differentiable around  $\theta_0$ with gradient $\nabla_\theta \psxn^{(para)}(\theta_0) \neq 0$.        
    Then, the bootstrap  confidence interval for $P^{(para)}_{S,x_n}(\theta_0)$
    given by 
    \[ \big[P_{S,x_n}^{(para)}(\widehat{\theta})   - q^*_{1-\delta/2},   \ \ P_{S,x_n}^{(para)}(\widehat{\theta})   - q^*_{\delta/2} \big], \]
    is asymptotically of  level  $1-\delta$.
     Here, $ q^\ast_\alpha$ denotes the $\alpha$-quantile of the distribution of $L_n^\ast:=P_{S,x_n}^{(para)}(\widehat{\theta}^\ast)-P_{S,x_n}^{(para)}(\widehat{\theta})$.
\end{Theorem}

As the above theorem shows,  the bootstrap confidence interval for the parametric predictive probability $ P_{S,x_n}^{(para)}(\theta_0) $ retains (asymptotically) the desired level $1-\delta$ even if the time series observed does not stem from the model class used to perform the prediction. As we will see in the simulations in Section \ref{sec:sims}, in this case  of model misspecification, the confidence intervals obtained inherit this uncertainty, which manifests itself  in larger but more honest   confidence intervals  compared to those obtained for the case where  $X_1, \ldots, X_n$ stems from the model class  ${\mathcal M}_\theta$ and this class also  is used to implement the bootstrap.

\subsubsection{Non-parametric Prediction} \label{subsec:bs_np}
Consider next  the case where  the  non-parametric estimator of the predictive probability $\widehat{P}_{S,x_n}^{(npara)}$ is used to perform the prediction; see the first row of Table~\ref{tab:T2}. In this case, the time series $ X_1, \ldots, X_n$ may or may not stem from the parametric family $ {\mathcal M}_\theta$. According to which one of the two possible scenarios should be  imitated in the bootstrap world, one can again use the estimated parametric one step transition probabilities, $P^{(para)}_{X_{t+1}=x_j|X_t=x_i}(\widehat\theta)$,   or the non-parametric analogue, 
$\widehat P^{(npara)}_{X_{t+1}=x_j|X_t=x_i}$,  to generate the bootstrap pseudo time series $ X_1^\ast, \ldots, X_n^\ast$.
 The following theorem can then be established.  

\begin{Theorem}[Non-parametric bootstrap confidence interval for $P_{S, x_n}=\psxn^{(npara)}$] \ \label{th.2.7}
\begin{enumerate} 
    \item[(i)] Suppose Assumptions \ref{assum_np} holds true and that $ X_1^\ast, \ldots, X_n^\ast$,  is generated using the non-parametrically estimated one step transition probabilities $ \widehat P^{(npara)}_{X_{t+1}=x_j|X_t=x_i}$.    Then, the bootstrap  confidence interval for $P_{S,x_n}$
    given by 
    \[ \big[\widehat{P}_{S,x_n}^{(npara)}   - q^*_{1-\delta/2},   \ \ \widehat{P}_{S,x_n}^{(npara)}   - q^*_{\delta/2} \big], \]
    is asymptotically of  level  $1-\delta$.
    Here, $ q^\ast_\alpha$ denotes the $\alpha$-quantile of the distribution of $L_n^\ast:=\widehat{P}_{S,x_n}^{\ast,(npara)}-\widehat{P}_{S,x_n}^{(npara)}$.
    \item[(ii)] Suppose Assumptions \ref{assum_np}, \ref{assum_p} and \ref{asum_p_bs} hold true and that $ X_1^\ast, \ldots, X_n^\ast$ is generated using the estimated, model-based one step transition probabilities $P^{(para)}_{X_{t+1}=x_j|X_t=x_i}(\widehat\theta)$. Then, the confidence interval constructed as in (i) is asymptotically of  level  $1-\delta$.
    \end{enumerate}
\end{Theorem}

\bigskip

\section{Parametric Prediction Using Specific Estimators and Models} \label{sec:inar1}
We illustrate our proposed prediction methodology  by an application to the  case where the popular INAR(1) and  INARCH(1)  models for count time series are used for prediction.  We also concentrate on  (conditional) maximum likelihood estimators of the parameters of these  models.
We first elaborate on properties of conditional maximum likelihood estimators under a setting which allows for model misspecification. 

\subsection{Conditional Maximum Likelihood Estimators} \label{subsec:cmle}

Let  
\[ l_n(\theta|X_1)=\sum_{t=2}^n \log\big(P^{(para)}_{X_t|X_{t-1}}(\theta)\big)\] 
be the (conditional on $X_1$) log-likelihood function, where 
 $ P^{(para)}_{X_t|X_{t-1}}(\theta)$  denotes  the one step transition probabilities of the process $(X_t(\theta),t\in\Z)$  belonging to the class $ {\mathcal M}_\theta$.  We denote by  $ \widehat{\theta}_{ML} =\mbox{argmax}_{\theta\in\Theta}l_n(\theta|X_1)$,
the (conditional) maximum likelihood estimator of $ \theta$ based on the time series $X_1, \ldots, X_n$ stemming from a  process $ (X_t, t \in \Z)$ satisfying Assumption \ref{assum_np}. 

Since we are in the setting of a parametric prediction and we will use the estimator  $\psxn^{(para)}(\widehat\theta_{\text{ML}}) $ of the predictive probability, in order  to construct a confidence interval for  $\psxn^{(para)}(\theta) $,   a parametric or a non-parametric bootstrap approach to generate the bootstrap time series $ X_1^\ast, \ldots, X_n^\ast$ can be used; see  the second row of Table \ref{tab:T2}. Recall that  using a non-parametric  approach in this context to generate the pseudo time series,  enables us to imitate  the distribution of $ \psxn^{(para)}(\widehat{\theta}_{ML})$ under model misspecification.   To establish validity of the bootstrap in this setting, the key step is to show that  the estimator $ \widehat{\theta}_{ML}^\ast$ satisfies Assumption \ref{asum_p_bs}.  
Towards this goal, some additional assumptions regarding the properties of the underlying process have to be imposed. We begin with the following assumption. 

\begin{asum}[Data generating process  under potential model misspecification] \label{assum_ML1} {~}
\begin{enumerate}
\item[(i)]  $ {\rm E}\big(\log P^{(para)}_{{X_{t+1}|X_{t}}}(\theta)  \big) $
has a unique maximum at $\theta_0$ in the interior of  $ \Theta$.
\item[(ii)] $ P^{(para)}_{X_{t+1}|X_{t}}(\theta)$ is two times differentiable around $\theta_0$  with respect to $\theta$ with Lipschitz continuous second-order partial derivatives. 
\item[(iii)] $ {\rm E}\|\nabla_\theta \log P^{(para)}_{X_{t+1}|X_{t}}(\theta) \|^2<\infty $ for every $ \theta\in\Theta$ and  the matrix  $ {\rm E}\big(\nabla^2_\theta \log P^{(para)}_{X_{t+1}|X_{t}}(\theta_0)\big) $ is positive definite.
\item[(iv)] ${\rm E}\big( \nabla_\theta \log P^{(para)}_{X_{t+1}|X_t}(\theta) \big) = \nabla_\theta {\rm E}\big(\log P^{(para)}_{X_{t+1}|X_t}(\theta) \big) $ for every $ \theta \in\Theta$.
\end{enumerate}
\end{asum}

  Part (i) of this assumption guarantees the uniqueness  of the limit of  the  maximum likelihood estimator which is important   when $(X_t,t\in\Z)$ does not  necessarily belong to $ {\mathcal M}_\theta$.   
 Assumption \ref{assum_ML1}(ii) imposes smoothness  conditions on the  parametric one step transition probabilities, Assumption \ref{assum_ML1}(iii) ensures the existence of second-order moments while  Assumption \ref{assum_ML1}(iv) allows for the interchangeability of expectation and differentiation. Such assumptions are   common for investigating  the consistency  and  distributional properties  of maximum likelihood  estimators; see for instance Condition 1.1 in \citet{billingsley61}.

Using  Assumption~\ref{assum_ML1}(ii),   a  Taylor series expansion of $ l_n^{'}(\theta|X_1)=\sum_{t=2}^n \nabla _\theta \log(P^{(para)}_{X_t|X_{t-1}}(\theta) ) $ around $ \theta_0$ and  the fact that $ l^{'}_n(\widehat{\theta}_{ML}|X_1)=0$,  we get   the basic expression  
\begin{align} \label{eq.ML-exp1} 
\frac{1}{n-1}l^{''}_n(\widetilde{\theta}_n|X_1) \sqrt{n-1}\big(\widehat{\theta}_{ML}-\theta_0)\big)
&  = -\frac{1}{\sqrt{n-1}}  \sum_{t=2}^n \nabla_\theta \log P^{(para)}_{X_t|X_{t-1}}(\theta)\big|_{\theta=\theta_0},
\end{align}
for some $ \widetilde{\theta}_n\in \Theta$ such that  $ \|\widetilde{\theta}_n-\theta_0\| \leq \|\widehat{\theta}_{ML}-\theta_0\|$.  We first establish the following result which  generalizes consistency of the (conditional) maximum likelihood estimator also for  case where the time series at hand does not necessarily stem from the model class $ {\mathcal M}_\theta$.

\begin{Theorem}[Convergence of $\widehat{\theta}_{ML}$] \label{th.ConML} Assume that $X_1,  \ldots, X_n$ stems from a process satisfying Assumption \ref{assum_np}. Let $ {\mathcal M}_\theta$ be a model class satisfying Assumption \ref{assum_p}(i) and suppose that  Assumption \ref{assum_ML1}(i) is fulfilled. Then, as $n\rightarrow \infty$, 
\[ \widehat{\theta}_{ML} \stackrel{P}{\rightarrow} \theta_0.\]
\end{Theorem}

Note that   analogous to (\ref{eq.ML-exp1}), a similar expression can also be obtained  in the bootstrap world  based on the pseudo time series $X_1^\ast,  \ldots, X_n^\ast$, that is,  we have,
\begin{align} \label{eq.ML-exp1-Boot} 
\frac{1}{n-1}l^{''}_n(\widetilde{\theta}^\ast_n|X^\ast_1) \sqrt{n-1}\big(\widehat{\theta}^\ast_{ML}-\widehat{\theta}_{ML}\big)
&  = -\frac{1}{\sqrt{n-1}}  \sum_{t=2}^n \nabla_\theta \log (P^{(para)}_{X^\ast_t|X^\ast_{t-1}}(\theta))\big|_{\theta=\widehat{\theta}_{ML}},
\end{align}
where $ \|\widetilde{\theta}^\ast_n-\widehat{\theta}_{ML} \| \leq \|\widehat{\theta}^\ast_{ML}-\widehat{\theta}_{ML}\|$.

Now,  
 when a parametric  model from the class ${\mathcal M}_\theta$ is fitted using  (conditional) maximum likelihood estimation and this model is used in the bootstrap procedure to generate the pseudo time series $X_1^\ast, \ldots, X_n^\ast$, then    the generated  bootstrap   data   stems from  a parametric model with parameter vector $\widehat{\theta}_{ML}$ irrespective of whether this also holds true  or not  for the observed time series $X_1, \ldots, X_n$.  In this case, and as we will see in the following,   asymptotic normality of the bootstrap sequence
\begin{equation}  \label{eq.BootSeq1}
\frac{1}{\sqrt{n-1}} \sum_{t=2}^n \nabla_\theta \log(P^{(para)}_{X^\ast_t|X^\ast_{t-1}}(\theta)\big)\Big|_{\theta=\widehat{\theta}_{ML}}
\end{equation}
can  typically  be established using a central limit theorem for triangular arrays of martingale differences; see expression (\ref{eq.MD-Array}) in the Appendix.  However, in the case where $X_1^\ast, \ldots, X_n^\ast$ is generated non-parametrically  using the estimated one step transition probabilities $\widehat P^{(npara)}_{X_{t+1}=x_j|X_t=x_i}$,  such a  property 
for the corresponding bootstrap sequence does not necessarily hold true. 
 In this case,  the asymptotic normality of 
the bootstrap  sequence (\ref{eq.BootSeq1})  relies on mixing properties which are related to corresponding properties 
of the underlying process $ (X_t,t\in\Z)$. To elaborate, consider the sequence 
\begin{align} \label{eq.ForCLT}
\frac{1}{\sqrt{n-1}} \sum_{t=2}^n & \nabla_\theta \log(P^{(para)}_{X_t|X_{t-1}}(\theta_0)\big) \nonumber \\
& =\frac{1}{\sqrt{n-1}} \sum_{t=2}^n \Big(\nabla_\theta \log(P^{(para)}_{X_t|X_{t-1}}(\theta_0)\big) -  {\rm E}\nabla_\theta \log(P^{(para)}_{X_t|X_{t-1}}(\theta_0)\big)\Big), 
\end{align}
appearing in   (\ref{eq.ML-exp1}) and where ${\rm E}(\nabla_\theta \log(P^{(para)}_{X_t|X_{t-1}}(\theta_0)\big))=0$  by Assumption \ref{assum_ML1}(iv). 
For instance, it is well known that countable Markov chains satisfying Assumption \ref{assum_np} are   $\beta$-mixing with  $\beta(k)\rightarrow 0$ as $ k\rightarrow \infty$; see \citet{bradley07}, Theorem 7.7.  The maximal correlation coefficient ($ \rho$-mixing), to which we focus  in the following,  is an alternative  way to control the dependence structure  of a Markov process and  to  establish the  weak convergence result
\[ \frac{1}{\sqrt{n-1}} \sum_{t=2}^n  \nabla_\theta \log(P^{(para)}_{X_t|X_{t-1}}(\theta)\Big|_{\theta=\theta_0}
\stackrel{d}{\rightarrow} {\mathcal N}\Big( 0, {\rm E}\big(\nabla_\theta^2 \log(P^{(para)}_{X_t|X_{t-1}}(\theta_0)\big)\big) \Big).\]  
From this, Assumption~\ref{assum_ML1}, Theorem~\ref{th.ConML} and  expression (\ref{eq.ML-exp1}), we get  that 
 \begin{equation} \label{eq.Distr-ML}
  \sqrt{n}(\widehat{\theta}_{ML} - \theta_{0}) \overset{d}{\rightarrow} \mathcal{N}(0, V_{\theta_0}),
  \end{equation}
where $V^{-1}_{\theta_0}={\rm E}\big(\nabla_\theta^2 \log(P^{(para)}_{X_t|X_{t-1}}(\theta_0)\big)\big)$. We can now establish the following  important result for bootstrap consistency, which shows that the (conditional) maximum likelihood estimator fulfills  the requirements of Assumption~\ref{asum_p_bs}.

\begin{Theorem}[Asymptotic normality of $\widehat{\theta}_{ML}$] \label{th.Boot-ML-CLT}
    Suppose Assumptions \ref{assum_np} and  \ref{assum_p}(i)  and \ref{assum_ML1}  hold true. 
    \begin{enumerate}
\item[(i)] Let  the pseudo time series $ X_1^\ast,  \ldots, X_n^\ast$ be  generated using the  one step transition probabilities $P^{(para)}_{X_{t+1}=x_j|X_t=x_i}(\widehat\theta_{\text{ML}})$ for $ x_i,x_j\in range(X_0)$.
Then,  as $ n \rightarrow \infty$, 
  \begin{equation} \label{eq.BooML}
  \sqrt{n}(\widehat{\theta}^*_{ML} - \widehat\theta_{ML}) \overset{d}{\rightarrow} \mathcal{N}(0, V_{\theta_0}),
  \end{equation}
    in probability, where $V^{-1}_{\theta_0}={\rm E}_{\theta_0}\big(\nabla_\theta^2 \log(P^{(para)}_{X_t|X_{t-1}}(\theta)\big)\big|_{\theta=\theta_0}\big)$. 
    \item[(ii)] Let   the pseudo time series $ X_1^\ast,  \ldots, X_n^\ast$ be  generated using the non-parametrically estimated  one step transition probabilities $\widehat P^{(npara)}_{X_{t+1}=x_j|X_t=x_i}$.
   Assume that  $(X_t,t\in\mathbb{Z})$ is $\rho$-mixing with $\rho$-mixing coefficient $\rho_1< 1$ and $E(|X_0|^{12+\delta})<\infty$ for some $\delta>0$ and   
   \begin{align}\label{assumption_theorem_3.2(ii)}
\max_{1\leq t\leq n} \nabla_\theta \log(P^{(para)}_{X^\ast_t|X^\ast_{t-1}}(\theta) \big)\big|_{\theta=\widehat \theta_{ML}} = o_{P^*}(\sqrt{n}).
\end{align}
Then, 
   as $ n \rightarrow \infty$, assertion (\ref{eq.BooML}) holds true with $ V^{-1}_{\theta_0}={\rm E}\big(\nabla_\theta^2 \log(P^{(para)}_{X_t|X_{t-1}}(\theta)\big)\big|_{\theta=\theta_0}\big)$.
    \end{enumerate}
\end{Theorem}

\begin{Remark}
Note that in assertion (i) of the above theorem, the expectation denoted by  ${\rm E}_{\theta_0}(\cdot)$ in the definition of $ V_{\theta_0}$ is taken with respect to the probability measure of the parametric model with parameter $\theta_0$. Furthermore, in assertion (ii) of the same theorem,  the  expectations appearing in the definition of $V_{\theta_0}$ and denoted by ${\rm E}(\cdot)$ is  taken with respect to the probability measure  of the underlying Markov process, which does not necessarily belong to the parametric class ${\mathcal M}_\theta$. Clearly, $ {\rm E}(\cdot)={\rm E}_{\theta_0}(\cdot)$ if $(X_t,t\in\Z)$  belongs to  $ {\mathcal M}_\theta$.
\end{Remark}

\begin{Remark}\label{remark_rho_mixing}
According to \citet{Roberts1997}, any stationary Markov chain that is geometrically ergodic and reversible is $\rho$-mixing with $\rho_1<1$, where 
\begin{align}
    \rho_k = \rho\big(\sigma(X_k),\sigma(X_0)\big), \quad   \rho_k\leq \rho_1^k.
\end{align}
The  maximal correlation coefficient $\rho(\cdot,\cdot)$ is defined by
\begin{align*}
    \rho(\mathcal{A},\mathcal{B}) = \sup_{f\in L_2(\mathcal{A})\;  ,g\in L_2(\mathcal{B})}|Corr(f,g)|,
\end{align*}
where $L_2(\mathcal{A})$ is the space of all random variables that are $\mathcal{A}$-measurable and square integrable. By Theorem 4.4(b1) of  \citet{bradley07}, for a process $(X_t,t\in\Z)$, we have 
\begin{align*}
    \rho_1 = \sup_{f,g}\left\{\frac{\big|E(f(X_i)g(X_{i-1}))-E(f(X_i))E(g(X_{i-1}))\big|}{\sqrt{E(f^2(X_i))}\sqrt{E(g^2(X_{i-1}))}}; \|f(X_i)\|_2<\infty,\; \|g(X_{i-1})\|_2<\infty\right\},
\end{align*}
where we used the notation $\|X\|_p=(E(X^p))^{1/p}$ for $p>1$.
\end{Remark}


\subsection{The   INAR(1) Model} \label{subsec:inar1} 
The INAR(1) model was first introduced by \citet{mck} and \citet{alosh} and extended to order $p$ by \citet{duli}. Recall that according to the   INAR($p$) model,   $X_t$ is generated  as
\begin{align} \label{eq:inarp}
X_t = \alpha_1 \circ X_{t-1} + \ldots + \alpha_p \circ X_{t-p} + \varepsilon_t, \quad t \in \mathbb{Z},
\end{align}
where $\varepsilon_t \overset{\text{i.i.d.}}{\sim} G$  with probability mass function $G(k), \, k \in \mathbb{N}_0=\mathbb{N}\cup \{0\}$ and  the vector of  coefficients $\alpha=(\alpha_1, \ldots, \alpha_p)^\top \in (0,1)^p$ fulfills $\sum_{i=1}^p \alpha_i < 1$. Recall that in order to  account for the integer nature of the data, model (\ref{eq:inarp})  uses the binomial thinning operator \enquote{$\circ$} first introduced by \citet{steutel} and  defined as  
\begin{align} \label{eq:thinning}
\alpha_i \circ X_{t-i} = \sum\limits_{j=1}^{X_{t-i}} Z_j^{(t,i)}, 
\end{align}
with $\big(Z_j^{(t,i)}, j \in \mathbb{N}, t \in \mathbb{Z}\big), i \in \{1, \ldots,p\}$ being mutually independent  random variables with $Z_j^{(t,i)} \sim \text{Bin}(1, \alpha_i)$ and  independent of the innovation process ($\varepsilon_t, \, t \in \mathbb{Z}$). Here, $ \text{Bin}(L,p)$ denotes the binomial distribution with parameters $ L\in \mathbb{N}$ and $ p\in[0,1]$. Note that according to this construction, $\alpha_i \circ X_{t-i}|X_{t-i}\sim \text{Bin}(X_{t-i}, \alpha_i)$ holds. 

In  the case of an INAR(1) model,  
the one step transition probabilities are given by  
\begin{align} \label{eq:trans_prob_p1}
P^{(para)}_{X_t=x_{t}|X_{t-1}=x_{t-1}}(\theta)  = \sum\limits_{j=0}^{\min(x_t, x_{t-1})} \binom{x_{t-1}}{j} \alpha_1^j (1-\alpha_1)^{x_{t-1}-j} G(x_t-j)
\end{align}
for $x_t,x_{t-1} \in \N_0$. 
Assume now  that an INAR(1) model  with  specified innovation distribution $G_{\gamma}$ depending on a  parameter $\gamma$ is used for prediction. Letting   $\theta=(\alpha, \gamma)$, the predictive probability of this model  is denoted  by 
 $\psxn^{(para)}(\theta) =: \psxn^{(inar)}(\theta)$. By  \eqref{eq:trans_prob_p1} and for  the  (conditional) maximum likelihood estimator $\widehat\theta_{\text{ML}} = (\widehat\alpha_{\text{ML}},\widehat\gamma_{\text{ML}})$    of  $\theta$,  the predictive probability can be estimated 
 by   
\begin{align} \label{eq:phat_p}
 \psxn^{(inar)}(\widehat\theta_{\text{ML}}) = \sum\limits_{j \in S} \sum\limits_{k=0}^{\min(x_n,j)}\binom{x_n}{k} \widehat\alpha_{\text{ML}}^k(1-\widehat\alpha_{\text{ML}})^{x_n-k}G_{\widehat\gamma_{\text{ML}}} (j-k).
\end{align}
To obtain a  confidence interval for $ \psxn^{(inar)}(\theta_{0}) $  one can use  the 
limiting distribution  of $ \widehat{\theta}_{ML}$. 
To elaborate, it is well known that under appropriate conditions including the assumption that $ X_1,  \ldots, X_n$ stems from a INAR(1) model,  it holds true that, as $n\rightarrow \infty$, 
\begin{align} \label{eq:clt_p}
\sqrt{n} \, (\widehat{\theta}_{\text{ML}} - \theta_0) \overset{d}{\rightarrow} \mathcal{N}(0, I^{-1}(\theta_0)),
\end{align}
where $\theta_0 = (\alpha_0, \gamma_0)$ is the true parameter and $I(\theta_0)={\rm E}(J_t(\theta_0))$, where 
$J_t(\theta_0) $ is the Hessian of $ -\log P^{(inar)}_{X_{t}|X_{t-1}}(\theta)) $ evaluated at $ \theta_0$;  see  for instance \citet{billingsley61}, \citet{freeland_diss} and  \citet{bookweiss}. Note that   $ {\rm E}(J_t(\theta_0)) $ can be estimated by  $(n-1)^{-1} \sum\limits_{t=2}^n J_t(\widehat{\theta}_{ML})$; also  see Remark B.2.1.2 in \citet{bookweiss}.

\begin{Remark} 
The asymptotic result   \eqref{eq:clt_p} holds true  for  a broad variety of INAR processes including  the prominent examples of a  Poisson INAR(1), (Poi-INAR(1)), and a negative binomial INAR(1), (NB-INAR(1)), process. 
\end{Remark}

From    \eqref{eq:clt_p}, we immediately get 
\begin{align} \label{eq:clt_p_delta}
\sqrt{n-1} \, ( \psxn^{(inar)}(\widehat\theta_{\text{ML}}) - \psxn^{(inar)} (\theta_0)) \overset{d}{\rightarrow} \mathcal{N}(0, s_{\theta_0}^2),
\end{align}
where $s^2_{\theta_0} = \nabla_\theta g(\theta_0)^\top I^{-1}( \theta_0) \nabla_\theta g(\theta_0)$ and $ \nabla_\theta g(\theta_0) = \sum_{j\in S}\nabla_\theta P^{(inar)}_{X_{t+1=j}|X_t=x_n}(\theta)\big|_{\theta=\theta_0}$. This  leads to the  asymptotically valid $(1-\delta)$-confidence interval  
\begin{align*}  \label{eq.ci-INAR1}
\left[ \psxn^{(inar)}(\widehat\theta_{ML}) - \frac{z_{\delta/2} \widehat s_{\theta_0}}{\sqrt{n-1}}, \ \psxn^{(inar)}(\widehat\theta_{ML}) + \frac{z_{\delta/2} \widehat s_{\theta_0}}{\sqrt{n-1}}  \right]
\end{align*}
for $ \psxn^{(inar)} (\theta_0)$,  where  $\widehat{s}^2_{\theta_0}=\nabla_\theta g(\widehat\theta_{\text{ML}})^\top I^{-1}(\widehat\theta_{\text{ML}}) \nabla_\theta g(\widehat\theta_{\text{ML}})$ and $z_{\delta/2}$ the $\delta/2$-quantile of the standard Gaussian distribution.

\begin{Remark} \label{rem_ex_g}
Note that in the case of a Poi-INAR(1) process, we have  for $\theta=(\alpha,\lambda)$, 
\[ g(\alpha,\lambda) := P_{S,x_n}^{(inar)}(\theta) = \sum\limits_{j \in S} \sum\limits_{k=0}^{\min(x_n,j)}\binom{x_n}{k} \alpha^k(1-\alpha)^{x_n-k}  \frac{\lambda^{j-k} e^{-\lambda}}{(j-k)!}. \] In the case of a NB-INAR(1) process, we have for $\theta=(\alpha,N,\pi)$, 
\[g(\alpha,N,\pi):=  P_{S,x_n}^{(inar)}(\theta)= \sum\limits_{j \in S} \sum\limits_{k=0}^{\min(x_n,j)}\binom{x_n}{k} \alpha^k(1-\alpha)^{x_n-k} \frac{\Gamma(j-k+N)}{\Gamma(N) (j-k)!} \pi^N (1-\pi)^{j-k}, \] where $\Gamma(\cdot)$ denotes the gamma function. 
For the Poi-INAR(1) case, we exemplarily get 
\begin{align} \label{eq:nabla_g_param}
\nabla g(\alpha, \lambda)) = \begin{pmatrix}
\sum\limits_{j \in S} \sum \limits_{k=0}^{\min(x_n,j)} \binom{x_n}{k} \frac{\lambda^{j-k}}{(j-k)!} e^{-\lambda} \alpha^{k-1} (k-x_n\alpha)(1-\alpha)^{x_n-k-1} \\
\sum\limits_{j \in S} \sum \limits_{k=0}^{\min(x_n,j)} \binom{x_n}{k} \alpha^k (1-\alpha)^{x_n-k} \frac{1}{(j-k)!} (-e^{-\lambda}) \lambda^{j-k-1} (-j+k+\lambda).
\end{pmatrix}
\end{align}
These expressions  have also  been used  for the numerical calculations in Section \ref{sec:sims}.
\end{Remark}

However, if  $ X_1, \ldots, X_n$ does \emph{not} stem from an INAR(1) model, but this  model is fitted to the data at hand, then one can take advantage of the convergence \eqref{eq.Distr-ML} 
in order to construct a confidence interval for the predictive probability $ \psxn^{(inar)}(\theta_{0})$ and 
proceed analogous to the case where (\ref{eq:clt_p}) holds true.
 The main difference in this case is  that 
$ V_{\theta_0} \neq I^{-1}(\theta_0)$ and   $ \theta_0=(\alpha_0, \gamma_0)$ is the unique parameter from the parameter space $\Theta$  which minimizes 
\[   {\rm E}\big(\log P^{(inar)}_{{X_t|X_{t-1}}}(\theta)  \big)=\sum_{r=0}^\infty\sum_{s=0}^\infty \log( P^{(inar)}_{X_t=r|X_{t-1}=s}(\theta)) P(X_t=r,X_{t-1}=s); \]
see Assumption \ref{assum_ML1}(i).

Instead of  using the asymptotic Gaussian distributions (\ref{eq:clt_p}), respectively, (\ref{eq.Distr-ML})  and in order to avoid estimation of  the  quantities $I^{-1}(\theta_0)$, respectively, $ V_{\theta_0}$,  one can use the bootstrap in both versions described  in the second row of Table~\ref{tab:T2} in order to obtain a confidence interval for $ \psxn^{(inar)}(\theta_{0})$.
  In view of the results obtained in Section \ref{subsec:cmle} and, in particular,  Theorem~\ref{th.Boot-ML-CLT}, the following  bootstrap confidence intervals can be constructed depending on whether one wants to imitate the situation of a time series stemming from the fitted model class or not. 

 Suppose that $X_1,  \ldots, X_n$ does not necessarily stem from an INAR(1) model, but that solely Assumption~\ref{assum_np} holds true. Let  the bootstrap pseudo time series $ X_1^\ast, \ldots, X_n^\ast$ be  generated using the  one step transition probabilities of the estimated INAR(1) model given  by   
\begin{align} \label{eq:trans_prob_p2}
P^{(inar)}_{X_{t}=x_t|X_{t-1}=x_{t-1}}(\widehat{\theta}_{ML})  = \sum\limits_{j=0}^{\min(x_t, x_{t-1})} \binom{x_{t-1}}{j} \widehat{\alpha}_{ML}^j (1-\widehat{\alpha}_{ML})^{x_{t-1}-j} f_{\varepsilon,\widehat{\gamma}_{ML}} (x_t-j).
\end{align}
 
Then,
the  bootstrap  confidence interval, 
   \begin{equation}
       \label{eq.Boot-CI-INAR1}
    \big[P_{S,x_n}^{(inar)}(\widehat{\theta}_{ML})  - q^*_{1-\delta/2},   \ \ P_{S,x_n}^{(inar)}{(\widehat{\theta}_{ML})}   - q^*_{\delta/2} \big], 
   \end{equation}
    is asymptotically of  level  $1-\delta$ for $ P_{S,x_n}(\theta_0)$, where $ q^\ast_\alpha$ denotes the $\alpha$-quantile of the distribution of  $L^\ast_n=P^{\ast,(inar)}_{S,x_n}(\widehat{\theta}^\ast_{ML}) - P_{S,x_n}^{(inar)}(\widehat{\theta}_{ML})$.
    
   Alternatively and  if the conditions stated in Theorem~\ref{th.Boot-ML-CLT}(ii) hold true and  the bootstrap  pseudo time series $ X_1^\ast, \ldots, X_n^\ast$ is   generated using the non-parametrically estimated  one step transition probabilities $\widehat P^{(npara)}_{X_{t+1}=x_j|X_t=x_i}$, then,  using the corresponding bootstrap distribution, a confidence interval analogue to (\ref{eq.Boot-CI-INAR1}) can be constructed. 
%

\subsection{The INARCH(1) Model} \label{subsec:inarch1}

The INARCH model belongs to the class of more general INGARCH models, which have been introduced by \citet{heinen_ingarch} (under a different name) and further analyzed by, e.g., \citet{ferland_ingarch} and \citet{fok09}. The INARCH($p$) model is described as 
\begin{align} \label{eq:inarchp}
    X_t | X_{t-1}, X_{t-2}, \ldots \sim Poi(\beta + \alpha_1 X_{t-1} + \ldots + \alpha_p X_{t-p}),
\end{align}
where $\beta > 0$, $\alpha_1, \ldots, \alpha_p \geq 0$ and $\sum_{i=1}^p \alpha_i < 1$. The one step transition probabilities of  model \eqref{eq:inarchp} are given by  
\begin{align} \label{eq:trans_prob_inarchp}
    P^{(para)}_{X_t=x_t|X_{t-1}=x_{t-1}, \ldots, X_{t-p}=x_{t-p}} = \exp \left(-\beta- \sum\limits_{i=1}^p \alpha_i x_{t-i} \right) \frac{\left(\beta+ \sum\limits_{i=1}^p \alpha_i x_{t-i} \right)^{x_{t}}}{x_t!}.
\end{align}
For $p=1$, we get the special case of an INARCH(1) model, which has been discussed in  \citet{weiss_inarch1} and the 
corresponding transition probabilities reduce to
\begin{align} \label{eq:trans_prob_inarch1}
    P^{(para)}_{X_t=x_t|X_{t-1}=x_{t-1}}  = \exp(-\beta-\alpha x_{t-1} ) \frac{(\beta+\alpha x_{t-1})^{x_{t}}}{x_t!}.
\end{align}

Assuming that an INARCH(1) model with parameter $\theta=(\beta,\alpha)$ is  used  for prediction and using the maximum likelihood estimator $\widehat\theta_{\text{ML}}=(\widehat\beta_{\text{ML}}, \widehat\alpha_{\text{ML}})$, the estimated predictive probability $ \psxnp(\theta_0) =:  \psxn^{(inarch)}(\theta_0)$ is given by 
\begin{align}
     \psxn^{(inarch)} (\widehat\theta_{\text{ML}}) = \sum\limits_{j \in S}  \exp(- \widehat\beta_{\text{ML}}-\widehat\alpha_{\text{ML}} x_n ) \frac{(\widehat\beta_{\text{ML}}+\widehat\alpha_{\text{ML}} x_n)^{j}}{j!}.
\end{align}

To construct an asymptotic or a bootstrap confidence interval for 
$ \psxn^{(inarch)}(\theta_0)$, one can proceed as discussed in Section~\ref{subsec:inar1} for the case of an INAR(1) model. 

Notice that under the assumption that $(X_t,t\in\Z)$ is an INARCH(1) model, 
\citet{zhu_wang} proved that $
    \sqrt{n-1}(\widehat\theta_{\text{ML}} - \theta_0) \overset{d}{\rightarrow} \mathcal{N}(0, I(\theta_0)^{-1})$,
where   $ I(\theta_0)$ can be estimated  by $(n-1)^{-1} \sum \limits_{t=2}^n J_t(\theta_0)$ and  
\[  J_t(\theta_0) = \begin{pmatrix}
    \frac{X_t}{(\beta + \alpha X_{t-1})^2} & \frac{X_t X_{t-1}}{(\beta + \alpha X_{t-1})^2} \\
    \frac{X_t X_{t-1}}{(\beta + \alpha X_{t-1})^2} & \frac{X_t X_{t-1}^2}{(\beta + \alpha X_{t-1})^2}
\end{pmatrix}.   \]
From this  we get
\begin{align}
    \sqrt{n-1}( \psxn^{(inarch)} (\widehat\theta_{\text{ML}}) - \psxn^{(inarch)} (\theta_0)) \overset{d}{\rightarrow} \mathcal{N}(0,s^2),
\end{align}
where $s^2 = \nabla g(\theta_0)^\top I^{-1}( \theta_0) \nabla g(\theta_0)$ with $$g(\theta_0) = \sum \limits_{j \in S}\exp(-\beta-\alpha x_n) \frac{(\beta+\alpha x_n)^{x_n}}{j!}$$ 
and the gradient given by 
\begin{align} \label{eq:nabla_g_inarch}
 \nabla g(\beta,\alpha) = \begin{pmatrix}
   \sum\limits_{j \in S} \frac{1}{j!} \exp(-\beta-\alpha x_n)(\beta + \alpha x_n)^{j-1}(j-\beta-\alpha x_n) \\  \sum\limits_{j \in S} \frac{1}{j!} \exp(-\beta-\alpha x_n)x_n(\beta + \alpha x_n)^{j-1}(j-\beta-\alpha x_n)
\end{pmatrix}. 
\end{align}


\section{Practical Issues and Extensions} \label{sec:pie}

\subsection{Range Preserving Confidence Intervals} \label{subsec:prac_prob}
It may happen that the confidence intervals obtained  exceed the natural range of a probability measure, i.e., the lower bound of the confidence interval is less  than zero or the upper bound exceeds one. This situation is more likely to occur  when predictive   sets $S$ are considered for which  $P(X_{n+1}\in S|X_n=x_n)$ is very close to the boundaries of $[0,1]$. 
To enforce the intervals to preserve the proper range, one way is to  use  the percentile method.  This works  by arranging  the  $B$ bootstrap replicates $\widehat P_{S,x_n}^{*,(b)}$, $b=1,2, \ldots, B$,   in increasing order, $\widehat P_{S,x_n}^{*,(1)}, \ldots, \widehat P_{S,x_n,}^{ *, (B)}$,
and use $\widehat P_{S,x_n}^{*,(B \delta/2)}$ as the lower bound and $\widehat P_{S,x_n}^{*, (B(1-\delta/2))}$ as the upper bound of the confidence interval. 
In the case that   $B \delta/2$ is not an integer, we define   $m= \lfloor (B+1) \delta/2 \rfloor$ and use $\widehat P_{S,x_n}^{*, (m)}$ as lower and $\widehat P_{S,x_n}^{*, (B+1-m)}$ as upper bound of the confidence interval. 

\subsection{Larger Prediction Horizon and Higher Order Processes} \label{subsec:extension}

Consider the case where prediction is made for a horizon $h>1$. In this case, one can think of the prediction problem as that   of obtaining a point or an interval estimator of the predictive probability
\begin{equation} \label{eq.Pred-h}
P\big(X_{n+j} \in S_j, \ j=1,2, \ldots, h|X_{n}=x_n\big),
\end{equation}
where $ S_j\subset \mathbb{R}$, $j=1, \ldots, h$,   are user-selected sets of values that   the process can take.  \eqref{eq.Pred-h}  refers to the probability that for the  future time points 
$t=n+1, \ldots, n+h$,  the Markov process will visit consecutively  the sets $S_1, \ldots, S_h$. Clearly, $ S_j=S $ for all $j=1,\ldots, h$ as well as $ P(X_{n+h}\in S|X_n=x_n)$, that is, $ S_1=S_2=\cdots=S_{h-1}=range(X_0)$, are   special cases of   (\ref{eq.Pred-h}).

Due to the Markov property, the predictive probability (\ref{eq.Pred-h}) can be expressed as a sum of proper one step transition probabilities, that is,  
\begin{align*}
    P\big(& X_{n+j}  \in S_j, \ j=1,2,   \ldots, h|X_{n}=x_n\big)\\ &= \sum_{y_h\in S_h}\sum_{y_{h-1}\in S_{h-1}}\cdots \sum_{y_1\in S_1} P_{X_{n+h}=y_h|X_{n+h-1}=y_{h-1}} P_{X_{n+h-1}=y_{h-1}|X_{n+h-2}=y_{h-2}} \cdots P_{X_{n+1}=y_1|X_{n}=x_{n}}.    
\end{align*}
This implies that estimating parametrically or non-parametrically the one step transition probabilities $P_{X_{t+1}=y_k|X_{t}=y_{k-1}} $ for $k=1,\ldots, h$ and  $ y_0\equiv x_n$, which are involved in the above expression, a parametric or non-parametric estimator of  the predictive probability $ P\big( X_{n+j}  \in S_j,  j=1,2,   \ldots, h|X_{n}=x_n\big)$ can be obtained.  Furthermore, an  asymptotic or bootstrap confidence interval for the predictive probability  of interest can in principle be constructed following  the approaches discussed in the previous sections.

When considering Markov processes of higher order $p>1$, the goal is to construct a confidence interval for  
\[  P_{S,(x_n, \ldots, x_{n-p+1})} := P(X_{n+1} \in S |X_{n-p+1}=x_{n-p+1},\ldots,X_n=x_n). \] 
For this, the main  construction principle  remains   the same for both  the asymptotic and the bootstrap approaches discussed. 
Note that  a non-parametric  estimator of $P_{S,(x_n, \ldots, x_{n-p+1})}$ is given by 
\[\widehat P_{S,(x_n, \ldots, x_{n-p+1})}^{\text{(npara)}} = \frac{\sum \limits_{t=1}^{n-p} \mathbf{1}_{\{ X_{t+1} \in S, X_{t-p+1} = x_{n-p+1}, \ldots, X_t = x_n \}}}{\sum \limits_{t=1}^{n-p} \mathbf{1}_{\{X_{t-p+1}=x_{n-p+1}, \ldots,  X_t = x_n \}}}. \] Analogously, a parametric estimator of the form $P_{S,(x_n, \ldots, x_{n-p+1})}^{\text{(para)}}(\widehat \theta)$ for $P_{S,(x_n, \ldots, x_{n-p+1})}^{\text{(para)}}(\theta_0)$ can be for instance obtained  by using  a $p$th order  INAR or INARCH model; see Section~\ref{subsec:inar1} and Section~\ref{subsec:inarch1}.  

\section{Simulations} \label{sec:sims}
We investigate the small sample performance of the asymptotic and the bootstrap methods to construct confidence intervals for the predictive probability.  For the  bootstrap, we distinguish between the construction principle   of  Algorithm \ref{algo_bs} and its percentile versions as discussed in Section \ref{subsec:prac_prob}. We  cover non-parametric and parametric setups focusing on predictions using  INAR(1) and INARCH(1) data generating processes (DGPs), as  elaborated in Section \ref{sec:inar1}. To estimate the asymptotic variances involved in the  non-parametric setup, we use the R package \textit{stableGR} \citep{stablegr}. For the asymptotic approach in the parametric setup, we proceed as follows: In case an INAR(1) model is assumed,  we use the asymptotic approach described in Section \ref{subsec:inar1} by applying  \eqref{eq:nabla_g_param} and the approximated Hessian of the R function \textit{constrOptim} \citep{r}. If an   INARCH(1) model is assumed, we use the procedure described in Section~\ref{subsec:inarch1}.

We use $K=500$ Monte Carlo samples and $B=500$ bootstrap repetitions  for four different sample  sizes $n\in\{100,500,1000,5000\}$. The significance level $\delta$ is set equal  to $5\%$. We consider time series that stem from a Poi(1)-INAR(1)  and from a NB(2,2/3)-INAR(1) model, both with parameter $\alpha=0.5$ and from an INARCH(1) model  with  parameters $\alpha=0.5$ and $\beta=1$. If not stated otherwise, we consider  $S=\{1,2\}$. 
For simulation,  estimation and bootstrapping of the INAR data, we use the R package \textit{spINAR} \citep{faymonville2024spinar}. For the estimation of the INARCH data, we use the R package \textit{tscount} \citep{tscount}.

\subsection{Asymptotic Confidence Intervals} \label{subsec:sim_asy_ci}
We apply the asymptotic procedures introduced in Section~ \ref{subsec:main_p} and Section~\ref{subsec:main_np} to different data generating processes (DGPs) and construct confidence intervals for $\psxn$. We would like to emphasize that also in the parametric procedure introduced in Section \ref{subsec:main_p}, we construct a confidence interval for $\psxn$ only coinciding with $\psxn(\theta_0)$ in case the assumed model for prediction is correct. To construct a confidence interval  for $\psxn(\theta_0)$, we would require an estimator for $V_{\theta_0}$, see Proposition \ref{prop22}, which is not straightforward in general. To avoid the need of such an estimator, we explicitly refer to the bootstrap procedures in \ref{sec:bootstrap}. The results obtained are presented 
in Table~\ref{tab:poi_inar1}. First, consider the case where the DGP is a Poi(1)-INAR(1) process. Comparing  coverages and mean lengths of the asymptotic confidence intervals, we see that the parametric approach using  a Poi-INAR model for prediction performs best. The  non-parametric approach keeps up well with respect to coverage, but it  leads to much wider intervals. Furthermore and as expected, the  parametric approach incorrectly using  an INARCH model for prediction behaves worse and delivers confidence intervals the coverage of which   decreases 
as the sample size $n$ increases. 
Consider next the case where the data stems from an  INAR(1) process with innovations having  an  overdispersed NB(2,2/3) distribution. 
In this case, both parametric models  (Poi-INAR and INARCH) used for prediction  are wrong, and this fact manifests itself in low   coverages which decrease  as $n$  increases. 
The  non-parametric approach  performs  best and shows a  good  coverage. The last considered case is that  of  time series stemming from  an INARCH(1) process. Here again, using the wrong model for prediction leads to coverages which deteriorate as the sample size increases. At the same time and as expected, the parametric approach using the correct model  performs best. The non-parametric approach lead to good results as this was the case in both previously considered DGPs.

\begin{table}[t]
\centering
\begin{tabular}{llc|rrrr}
\toprule
 Data  &     Prediction    & & & &  &  \\
 & &$n$& $100$ & $500$ & $1000$ & $5000$ \\ 
\midrule
 Poi-INAR  &  Poi-INAR &cov & 0.934&0.942&0.952&0.958 \\ 
  &  & ml & 0.133&0.059&0.044&0.019 \\ 
  &  INARCH  & cov & 0.440&0.352&0.354&0.122 \\ 
 &  &ml & 0.101&0.043&0.034&0.014 \\ 
   & npara  & cov & 0.915&0.944&0.934&0.964 \\ 
 &  &ml & 0.423&0.199&0.151&0.065 \\
 \midrule 
 NB-INAR & Poi-INAR &cov &  0.716&0.550&0.484&0.294 \\ 
& & ml & 0.123&0.054&0.039&0.017\\ 
  & INARCH  & cov & 0.548&0.190&0.140&0.086 \\ 
 & &ml & 0.105&0.045&0.033&0.014 \\ 
  &  npara  & cov & 0.890&0.944&0.926&0.944 \\ 
  &&ml & 0.441&0.221&0.152&0.070 \\ 
\midrule 
   INARCH & Poi-INAR &cov &  0.732&0.302&0.174&0.084 \\ 
 && ml & 0.115&0.050&0.036&0.017 \\ 
& INARCH  & cov & 0.932&0.918&0.918&0.944 \\ 
 & &ml & 0.101&0.042&0.031&0.014\\
 &npara  & cov & 0.913&0.950&0.926&0.968 \\ 
 &&ml & 0.452&0.211&0.155&0.070 \\ 
 \bottomrule
\end{tabular}
\vspace{0.3cm}
\caption{Coverage (cov) and mean length (ml) of asymptotic confidence intervals for $\psxn$ for different sample sizes, different parametric and non-parametric approaches and significance level 0.05 when applied to different DGPs.  }
\label{tab:poi_inar1}
\end{table}

To summarize, we see that concerning the calculation of a confidence interval for $ P_{S,x_n}$, there is a trade-off between robustness and efficiency. The non-parametric approach is robust to model misspecification. Using a parametric approach is more efficient when the model  assumptions made hold true, but it can lead to worse coverages if this is not the case because then $ P_{S,x_n}(\widehat{\theta})$ is not a consistent  estimator of $ P_{S,x_n}$ .


\subsection{Bootstrap Confidence Intervals}

We next investigate the finite sample performance of  the bootstrap  for   constructing   confidence intervals. 
We consider three of the four possible scenarios described  in  Table~\ref{tab:T1}, respectively, Table~\ref{tab:T2}.
In the parametric case of \ref{subsec:bs_p}, we construct confidence intervals for the parametric predictive probability $\psxn^{(para)}(\theta_0)$. Recall that only if the  parametric model used is correct, the predictive probabilities $\psxn$ and $\psxn^{(para)}(\theta_0)$ coincide. 
In the non-parametric case of \ref{subsec:bs_np}, we construct confidence intervals for the non-parametric predictive probability coinciding with $\psxn$.

First of all, we investigate the parametric prediction setup  and, in particular, the two cases listed in  the second row of Table~\ref{tab:T1} and  Table~\ref{tab:T2}. For this purpose, a  Poi-INAR(1) model  is used for prediction  and the following two  scenarios for generating the bootstrap time series are  considered: The bootstrap time series is generated  non-parametrically, while 
the estimated Poi-INAR(1) model is used for prediction and 
the bootstrap time series is generated parametrically using the estimated  
Poi-INAR(1) model and this model also is used for prediction.
We generate time series  from the three different DGPs, that is, we use an INAR(1) process with $\alpha=0.5$ and Poi(1) innovations, an INAR(1) process with $\alpha=0.5$ and NB(2,2/3) innovations and an INARCH(1) process with $\alpha=0.5$ and $\beta=1$. Notice that all three processes possess  an expected value of  $2$. 
To accurately approximate the unknown probability  $\psxn^{(para)}(\theta_0)$, we use  a sample of $n=10^6$ observations. 

The results obtained are presented  in Table~\ref{tab:mms}. Observe that if the
DGP does not coincide with the Poi-INAR(1) used for prediction, the bootstrap delivers  good  coverages which are approaching the nominal $95\%$ for increasing $n$ in  the case where a  non-parametric bootstrap procedure is used to generate the pseudo time series.  However, when the bootstrap time series is generated  parametrically using the (wrong)  Poi-INAR model, the coverage is systematically lower. Also in terms of the mean length, there is a visible pattern if the model is misspecified. As the results show,  the confidence intervals obtained in this case by applying a  non-parametric bootstrap are always slightly larger than those obtained when  a parametric bootstrap is used. This reflects  the uncertainty associated with  the fact that 
 a wrong  model is used  to perform prediction. 
If the model is correctly specified, i.e., when we are in the case of a Poi-INAR DGP and the corresponding estimated  model is used for prediction, both bootstrap procedures work well with respect to both coverage and length. It seems that  in this 
case,  essentially no price is  paid  for generating the bootstrap time series non-parametrically, that is,  without using the (true) model.   In summary, the proposed procedure in the below right corner of Table \ref{tab:T2} is able to construct confidence intervals for $\psxn^{(para)}(\theta_0)$ that retain the desired coverage. 

\begin{table}[t]
\centering
\begin{tabular}{cccc|cccc}
\toprule
 Data  & Prediction &    Bootstrap    & & & &  &  \\
 & & &$n$& $100$ & $500$ & $1000$ & $5000$ \\ 
\midrule
NB-INAR & Poi-INAR & para & cov & 0.872 & 0.922 & 0.890   & 0.914    \\ 
 & & & ml & 0.128 & 0.057 & 0.041  &  0.018   \\ 
 & & npara & cov & 0.892 & 0.942 & 0.930 &   0.932    \\ 
 &&  & ml & 0.139 & 0.062 & 0.044  & 0.019   \\ 
\midrule
 INARCH & Poi-INAR & para & cov & 0.892 & 0.902 & 0.922  & 0.924    \\ 
  & & & ml & 0.122 & 0.053 & 0.039  & 0.018    \\ 
  & & npara & cov & 0.898 & 0.914 & 0.940  & 0.944    \\ 
  && & ml & 0.129 & 0.056 & 0.041  &  0.019   \\ 
\midrule
Poi-INAR & Poi-INAR & para & cov & 0.900 & 0.930 & 0.946  & 0.956    \\ 
  && & ml & 0.130 & 0.058 & 0.043   &  0.019   \\ 
  && npara & cov & 0.906 & 0.930 & 0.940 &   0.952    \\ 
&&  & ml & 0.132 & 0.058 & 0.043   &  0.019    \\ 
\bottomrule 
\end{tabular} 
\vspace{0.3cm}
\caption{Coverage (cov) and mean length (ml) of bootstrap  confidence intervals for the predictive probability $\psxn^{(para)}(\theta_0)$ when a  Poi-INAR(1) model is used for prediction, i.e., for $\psxn^{(inar)}(\lambda)$, where $\lambda = 1$ and $\lambda$ the parameter of the Poisson distribution. 
A parametric (para)  and a non-parametric  (npara) bootstrap  procedure is used and  applied to  three different DGPs and four different sample sizes.}
\label{tab:mms}
\end{table}

We next consider the case of non-parametric prediction combined with non-parametric bootstrapping, i.e., we are referring  to  the first row and second column of Table~\ref{tab:T2}.
In addition, we also compare the results with the corresponding percentile version described in Section~\ref{subsec:prac_prob}. Table~\ref{tab:mms_np} summarizes the results and shows  that for increasing $n$ and for all DGPs considered,  the coverage is good and improves  by getting  closer to  the desired 95\% level. While the mean length of the intervals is  large for small sample sizes, it  decrease  fast as $n$ increases. 
Similar results are obtained using the percentile version.  

\begin{table}[t]
\centering
\begin{small}
\begin{tabular}{lllr|rrrr|rrrr}
\toprule
        &&    &       & \multicolumn{4}{c|}{}&  \multicolumn{4}{c}{percentile version} \\
 Data& Prediction & Bootstrap &$n$& $100$ & $500$ & $1000$ & $5000$&$100$ & $500$ & $1000$ & $5000$ \\ 
\midrule
NB-INAR & npara & npara & cov & 0.894&0.944&0.922&0.938   &0.922&0.958&0.934&0.950 \\ 
 && &ml & 0.521&0.221&0.154&0.071   &0.526&0.222&0.156&0.072  \\
\midrule
INARCH & npara & npara  & cov &0.906&0.950&0.926&0.958    &0.930&0.960&0.936&0.968 \\ 
 &&&ml & 0.516&0.215&0.160 & 0.070   &0.532&0.222&0.164&0.072  \\
\midrule
Poi-INAR & npara & npara & cov & 0.922&0.940&0.922&0.954&0.938&0.956&0.936&0.958 \\ 
 &&&ml & 0.489&0.210&0.155&0.065&0.493&0.212&0.157&0.065 \\ 
 \bottomrule      
\end{tabular}
\end{small}
\vspace{0.3cm}
\caption{Coverage (cov) and mean length (ml) of bootstrap  confidence intervals for $\psxn$ for different sample sizes and three different true DGPs. The confidence intervals are obtained by  applying  Algorithm~\ref{algo_bs} with the steps specified as in the upper right corner of Table~\ref{tab:T2}.}
\label{tab:mms_np}
\end{table}

\subsection{Range-preserving Intervals}

Table \ref{tab:mms_np} displays the results when using the percentile version of the bootstrap confidence intervals. As pointed out in Section \ref{subsec:prac_prob},   the non-compliance with the natural range of probability can occur in situations where  the considered set $S$ is very (un)likely. To investigate  such a scenario, we reconsider the Poi-INAR(1) DGP with $\alpha=0.5$ and $\lambda=1$, but now we define $S=\{10\}$. In the case of a Poi-INAR(1) DGP, the Poisson distribution of the innovations transfers to the observations leading to a Poi($\lambda/(1-\alpha)$) marginal distribution, i.e., the observations are Poi(2) distributed. For $Y \sim \text{Poi}(2)$, we have $P(Y=10)\approx 3.8 \cdot 10^{-5}$ illustrating the \enquote{unlikeliness} of $S$. Table \ref{tab:param_s10} compares the results of the parametric asymptotic approach, the parametric bootstrap approach, where we use the parametric assumption for both estimating and bootstrapping, and its percentile version, where all three correctly assume a Poi-INAR(1) DGP. We see that the percentile intervals provide an increased and, consequently, more accurate coverage especially for small sample size $n$. This could be expected since we avoid the violation of the natural range of probabilities. Asymptotically, the approaches do not differ much.

\begin{table}[t]
\centering
\begin{small}
\begin{tabular}{lr|rrrr|rrrr}
\toprule
            &       & \multicolumn{4}{c|}{}&  \multicolumn{4}{c}{percentile version} \\
 &$n$& $100$ & $500$ & $1000$ & $5000$&$100$ & $500$ & $1000$ & $5000$ \\ 
\midrule
  Poi-INAR &cov &  0.854&0.930&0.928&0.946&-&-&-&- \\ 
  asy & ml ($\times 10^{-5}$) & $8.5$&$3.7$&$3.8$&$1.1$&-&-&-&- \\ 
\midrule
Poi-INAR  & cov & 0.750&0.840&0.884&0.936&0.962&0.954&0.946&0.948 \\ 
 bs &ml ($\times 10^{-5}$) & $9.6 $&$3.8$&$3.8$&$1.1$&$9.9$&$4.0$&$3.9 $&$1.1$ \\ 
 \bottomrule      
\end{tabular}
\end{small}
\vspace{0.2cm}
\caption{Coverage and mean length of the confidence intervals resulting from the three parametric approaches for different sample sizes and the different approaches applied on Poi(1)-INAR(1) DGP with $\alpha=0.5$ in case of $S=\{10\}$.}
\label{tab:param_s10}
\end{table}

\section{Real-Data Application} \label{sec:appl}
\subsection{Rotary Drilling Data}
\begin{figure}
\includegraphics[scale=0.33]{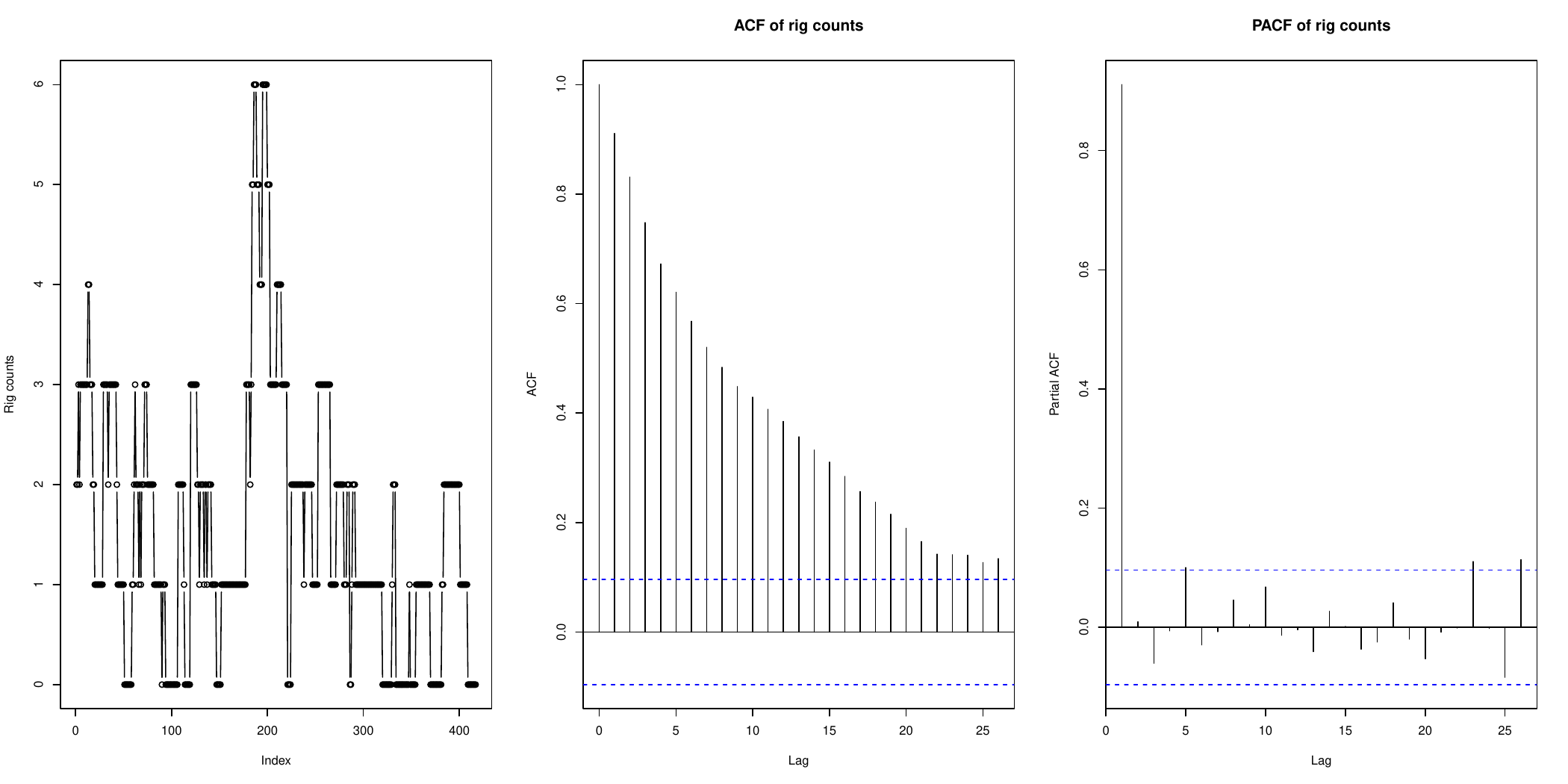}
\caption{Plot of time series of weekly active rotary drillings and its corresponding (P)ACF.}
\label{fig:rig}
\end{figure}

In a first illustration of our proposed predictive inference procedure, we consider a time series of length $n=417$ consisting of  weekly counts of active rotary drilling rigs in Alaska (1990-1997)\footnote{phx.corporate-ir.net/phoenix.zhtml?c=79687\&p=irol-rigcountsoverview}. They are used to measure the demand of products from the drilling industry and have previously been analyzed in \citet{bookweiss} and \citet{faymonville2024gof}. In Figure \ref{fig:rig}, we see that the time series contains a lot of runs, i.e., same values for consecutive time points, which indicates a strong serial dependence together with a small innovations' mean. The latter is supported by the high and slowly decreasing autocorrelation level  observed in the ACF plot. The PACF plot suggests a first-order  autoregressive structure with   $\widehat{\rho}(1) \approx 0.91$. An INAR(1) model appears to be reasonable. Indeed, testing the semi-parametric null hypothesis of an INAR(1) model, \citet{faymonville2024gof} were not able  to reject this hypothesis  at $5\%$ level. The data exhibits low counts and a dispersion index of approximately $1.11$ suggesting approximate equidispersed counts. 

Given that  $x_n=0$, a company in the drilling industry might be interested in the event that $x_{n+1}=0$, i.e., that also in the next week there are no active rotary drilling rigs. As a consequence, this will result in a low  demand of products of this company which may lead to a reduction of production activity. Apart from the  set  $S^{(1)}=\{0\}$, another set of interest could be the set $S^{(2)}=\{x: x \geq 2\}$. In this case, the  company may  want to hedge against suddenly more active rigs in the next week,  which is associated with a higher need of replacement parts.
In the following, we aim to  estimate the predictive probability of these two sets and to construct a confidence interval for $P_{S^{(i)}, x_n}, \, i=1,2,$ at $95\%$ level.
Since $x_n=0$ and taking into account the previous findings from the ACF and PACF plots, we may think of  the set  $S^{(2)}$ as 
 an unlikely event. We apply the parametric and non-parametric bootstrap method to construct confidence intervals  as described in Section~ \ref{subsec:bs_p} and Section~\ref{subsec:bs_np}. We also  additionally calculate the  percentile version of these confidence intervals; see Section~\ref{subsec:prac_prob}. In the parametric case, we assume a Poisson distribution for the innovations which seems to be reasonable  due to the equidispersed counts. The parametric assumption  of an INAR(1) is both used for the point estimation and  for the generation of the bootstrap time series. The same holds true also in the case where a  non-parametric approach is used. Table \ref{tab:rig_ci} shows the results of the predictive probability of the two sets considered.   As we see, the results for the parametric and the non-parametric procedure are close 
 to each other and the confidence intervals obtained for the non-parametric prediction are, as expected, wider than those obtained in the parametric case. Furthermore,  the percentile confidence intervals seem to be shorter and
 in the case of the set $S_2$, the benefits of these intervals  become clear. While the lower bounds of the two bootstrap confidence intervals fall below 0, the percentile approach respects the natural range of the underlying parameter. 

\begin{table}[ht]
\centering
\begin{tabular}{cc|ccc}
& & Predictive Probabilities & Bootstrap CI & Percentile Version \\
 \hline 
 $S^{(1)}$ &para& 0.8822  & [0.8402, 0.9631] &[0.8485, 0.9143] \\
 &npara&0.8764&[0.8309, 0.9681]&[0.7833,  0.9231] \\
 $S^{(2)}$ &para& 0.0072  & [-0.0067, 0.0115] &[0.0038, 0.0121] \\
 &npara&0.0337&[-0.0182, 0.0674]&[0.0000,  0.0909] 
\end{tabular}
\vspace{0.2cm}
\caption{Estimated predictive probabilities and (percentile) bootstrap-based confidence intervals for the 
sets $S^{(1)}=\{0\}$ and $ S^{(2)}=\{x:x\geq 2\}$ of the Rotary Drilling Data.}
\label{tab:rig_ci}
\end{table}

\subsection{Transaction Data} 
\begin{figure}[t]
\centering
\includegraphics[scale=0.33]{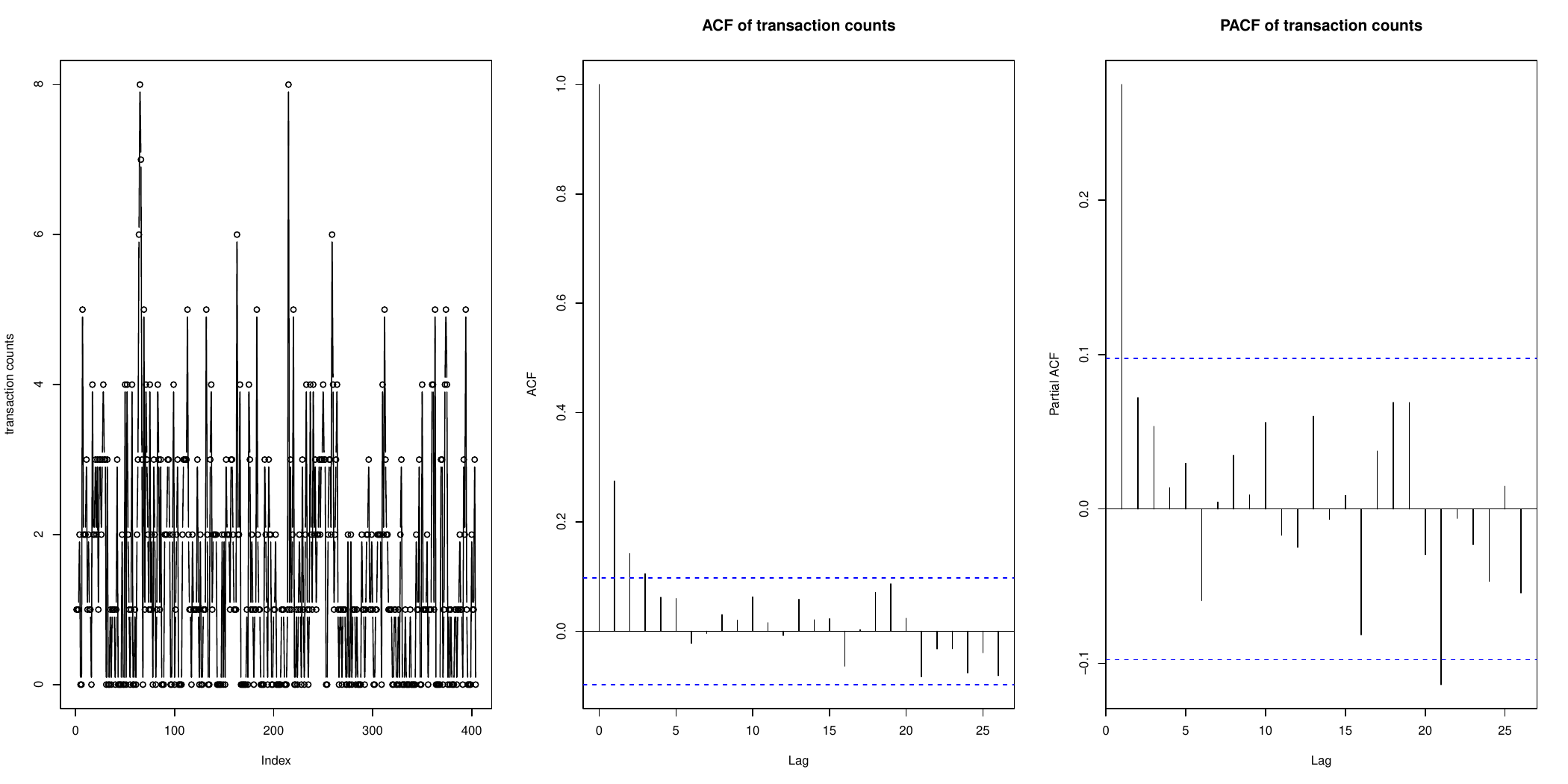}
\caption{Plot of time series of transaction counts and its corresponding (P)ACF.}
\label{fig:transact}
\end{figure}

In a  second application, we consider a data set provided by the Deutsche Börse Group which has been discussed  in \citet{data1} and \citet{faymonville2024gof}. For the period from February 2017 to  August 2019, this data set  contains $n=404$ counts of transactions of structured products per trading day. Figure \ref{fig:transact} shows a plot of the data along with the corresponding (P)ACF. Applying a goodness-of-fit test, \citet{faymonville2024gof} were not able to reject the semi-parametric null hypothesis of an INAR(1) model at $5\%$ level. 
Furthermore, they argued that  due to overdispersion,  a Poi-INAR(1) model would probably not be suitable for this  data set. Instead,  a Geo-INAR(1) model would be more appropriate  since the  semi-parametrically estimated probability function  of the innovations seems to be  rather  close to a geometric distribution. These findings motivate the application  of a  non-parametric and parametric prediction with bootstrap-based confidence intervals, where  for  obtaining these intervals,   non-parametric and parametric generation of the bootstrap time series is applied. We aim to construct a $95\%$ confidence interval for $\psxn$ and for  $\psxn(\theta_0)$ by  considering two different sets of interests,  $S^{(1)} = \{0\}$ and $S^{(2)}=\{2\}$. While the first one is associated with  a lack of interest in the respective financial product, the second set  speaks for a slight return to activity when $x_n=0$. Both sets of interests concern   questions related  to risk management and investment strategies.
In case of the parametric approaches, we separately consider  the case of  Poisson and geometrically distributed innovations. 
The results obtained are displayed in Table \ref{tab:transact_ci}.  Let us first discus these results   from  a goodness-of-fit point of view. By comparing the results of the parametric Po-INAR  approach with  those of the Geo-INAR and the non-parametric  approaches, we see that  the corresponding  confidence intervals are almost disjoint. Additionally, when  a geometric assumption for the innovations is used, the results seem more plausible and are very close to those obtained using a non-parametric method. In particular,  the confidence intervals obtained using the Geo-INAR(1) model,  are contained in the confidence intervals constructed using the non-parametric approach.  This is in line with  the findings  in  \citet{faymonville2024gof}, which argued in favor of  a geometric distribution for  the  innovations.  Let us spotlight the issue  of model misspecification in the context of this data example. Toward  this,  consider the rows 1-4 and 5-9 of Table~\ref{tab:transact_ci}. We see that in the Geo-INAR case, the confidence intervals based on the parametric and the non-parametric bootstrap are very similar. This suggests that the uncertainty about the correctness of the parametric model  associated with performing the prediction by using a Geo-INAR(1) model is almost   negligible. In other words, the Geo-INAR(1)   model describes this data set well, which seems not to be the case   for the Poi-INAR model. The confidence intervals obtained using  the latter model differ more to those obtained by the non-parametric method.

\begin{table}[ht]
\centering
\begin{tabular}{ccl|cc}
Set & Prediction  & Bootstrap & Predictive Probability& Bootstrap CI \\ 
\hline 
$S^{(1)}$ & Poi-INAR & para & 0.3203 & [0.2720, 0.3676] \\ 
 && npara &  & [0.2598, 0.3822] \\ 
&Geo-INAR & para & 0.4929 & [0.4469, 0.5290] \\ 
 && npara &  & [0.4549, 0.5245] \\ 
&npara & npara & 0.4884 & [0.4023, 0.5752] \\ 
\hline
$S^{(2)}$ & Poi-INAR & para & 0.2076 & [0.1841, 0.2343] \\ 
 && npara &  & [0.1822, 0.2372] \\ 
&Geo-INAR & para & 0.1267 & [0.1189, 0.1365] \\ 
 && npara &  & [0.1185, 0.1367] \\ 
&npara & npara & 0.1318 & [0.0676, 0.1824] \\ 
\end{tabular} 
\vspace{0.2cm}
\caption{Estimated predictive probabilities and bootstrap confidence intervals using different parametric and non-parametric approaches for the sets $S^{(1)}=\{0\}$ and $S^{(2)}=\{2\}$ of the Transaction Data.}
\label{tab:transact_ci}
\end{table}


\section{Conclusions} \label{sec:concl}

The construction of prediction intervals is a well-developed topic in the literature. However,  research is largely  limited to the setup of continuous-valued time series. 
For discrete-valued time series, classical prediction intervals are not meaningful while  prediction sets may  not be  able to keep a desired  coverage level. To solve this problem, we proposed to reserve  the  prediction  problem: Instead of constructing  prediction intervals or  sets given a desired coverage level, we argued to estimate  predictive probabilities  for  user-selected subsets of  values that the process can take in the future and to manage the uncertainty associated with the  prediction by constructing confidence intervals for the corresponding predictive probabilities. Different asymptotic and bootstrap-based procedures have been proposed for the construction of such confidence intervals, covering parametric and non-parametric setups and  allowing   to  address  the important case of possible model misspecification in implementing such a   prediction. 
  Whether a  parametric or a non-parametric approach is   used for prediction  is a decision that the user has to make.  The methodology proposed in this paper aims to provide a statistical way  to quantify the uncertainty associated with  the  prediction made. The corresponding  confidence intervals are constructed in such a way that  even    doubts  about the appropriateness  of the model used for prediction can properly be taken into account,  which is an important  issue  in case the decision of the user  is in favor of a parametric model. 


\section*{Acknowledgments}

The authors gratefully acknowledge the computing time provided on the Linux HPC cluster
at TU Dortmund University (LiDO3), partially funded in the course of the Large-Scale Equipment
Initiative by the German Research Foundation (DFG) as project 271512359.

The research of M. Faymonville and C. Jentsch was funded by the Deutsche Forschungsgemeinschaft (DFG, German Research Foundation) Project ID 437270842 (Model Diagnostics for Count Time Series) and 520388526 (TRR 391: Spatio-temporal Statistics for the Transition of Energy and Transport, Project A03). The research of  E. Paparoditis was also funded by  the  Cyprus Academy of Sciences,  Letters, and Arts.

\section*{References}
\markright{PREDICTIVE INFERENCE FOR DISCRETE-VALUED TIME SERIES}
\markleft{MAXIME FAYMONVILLE \& CARSTEN JENTSCH \& EFSTATHIOS PAPARODITIS}                     
\renewcommand*{\refname}{References}  
\markright{PREDICTIVE INFERENCE FOR DISCRETE-VALUED TIME SERIES}
\markleft{MAXIME FAYMONVILLE \& CARSTEN JENTSCH \& EFSTATHIOS PAPARODITIS}   
\printbibliography[title=References, heading = none]
\markright{PREDICTIVE INFERENCE FOR DISCRETE-VALUED TIME SERIES}
\markleft{MAXIME FAYMONVILLE \& CARSTEN JENTSCH \& EFSTATHIOS PAPARODITIS}   
\newpage 


\newpage 
\thispagestyle{empty} 

\appendix




\section{Auxiliary Results and Proofs}

\noindent{\bf Proof of Proposition~\ref{theo:ci_np}:} 
From \citet{derman1956}, we get that
\begin{equation}
\sqrt{n}\big(\widehat P^{(npara)}_{X_{t+1}\in S|X_t=x_n}  -P_{X_{t+1}\in S|X_t=x_n} \big)  \overset{d}{\rightarrow} \mathcal{N}\Big(0, \sigma_S^2\Big),
    \end{equation}
where $\sigma_S^2=a_S^T \Sigma_S a_S$. By making use of the Delta method and writing
\begin{align*}
\sqrt{n}\big(\widehat P^{(npara)}_{X_{t+1}\in S|X_t=x_n}  -P_{X_{t+1}\in S|X_t=x_n} \big) = & \sqrt{n}\left(\frac{\widehat Q_{S,x_n}}{\widehat Q_{x_n}} - \frac{Q
_{S,x_n}}{Q_{x_n}}\right)  \\
= & \sqrt{n}\left(g(\widehat Q_{S,x_n},\widehat Q_{x_n}) - g(Q_{S,x_n},Q_{x_n})\right),
\end{align*}
where $g(a,b)=a/b$ with $\nabla g(a,b) = (1/b,-a/b^2)^\top$ for $b>0$, the formula for the limiting variance $\sigma_S^2=a^T \Sigma a$ is easily obtained, where 
\begin{align*}
a_S = \nabla g(Q_{S,x_n},Q_{x_n}) = \left(\frac{1}{Q_{x_n}},-\frac{Q_{S,x_n}}{Q_{x_n}^2}\right)^\top
\end{align*}
and 
\begin{align}\label{eq:Sigma}
   \Sigma_S = \begin{pmatrix}
\Sigma_S^{(1,1)}  & \Sigma_S^{(1,2)}   \\
\Sigma_S^{(2,1)} & \Sigma_S^{(2,2)}
\end{pmatrix}
\end{align}
with $(\Sigma_S^{(2,1)})^\top=\Sigma_S^{(1,2)}$ and
\begin{align*}
   \Sigma_S^{(1,1)} = & \lim_{n\rightarrow \infty} nCov(\widehat Q_{S,x_n},\widehat Q_{S,x_n}) = \sum_{h=-\infty}^\infty Cov\left(\mathbf{1}_{ \{X_{h+1} \in S, X_{h}=x_n \}},\mathbf{1}_{ \{X_{1} \in S, X_{0}=x_n \}}\right)    \\
   = &\sum_{h=-\infty}^\infty \big( P(X_{h+1} \in S, X_{h}=x_n,X_{1} \in S, X_{0}=x_n)-P^2(X_{1} \in S, X_{0}=x_n)
   \big), \\
    \Sigma_S^{(1,2)} = & \lim_{n\rightarrow \infty} nCov(\widehat Q_{S,x_n},\widehat Q_{x_n}) = \sum_{h=-\infty}^\infty Cov\left(\mathbf{1}_{ \{X_{h+1} \in S, X_{h}=x_n \}},\mathbf{1}_{ \{ X_{0}=x_n \}}\right)    \\
   = &\sum_{h=-\infty}^\infty \big( P(X_{h+1} \in S, X_{h}=x_n,X_{0}=x_n)-P(X_{1} \in S, X_{0}=x_n) P(X_{0}=x_n) 
   \big), \\
    \Sigma_S^{(2,2)} = & \lim_{n\rightarrow \infty} nCov(\widehat Q_{x_n},\widehat Q_{x_n}) = \sum_{h=-\infty}^\infty Cov\left(\mathbf{1}_{ \{X_{h}=x_n \}},\mathbf{1}_{ \{ X_{0}=x_n \}}\right)    \\
   = &\sum_{h=-\infty}^\infty \big( P(X_{h}=x_n,X_{0}=x_n)-P^2(X_{0}=x_n) 
   \big),
\end{align*}
where the infinite sums in all three terms $\Sigma_S^{(1,1)}, \Sigma_S^{(1,2)}$ and $\Sigma_S^{(2,2)}$ converge due to the geometric ergodicity of the process.


\hfill $\Box$


\bigskip

\noindent {\bf Proof of Theorem~\ref{th.2.7}:} (i)
    Relying on  the estimated relative frequencies $\widehat{P}^{(npara)}_{X_{t+1}=j |X_{t}=i}$, the bootstrap procedure described in Step 2 of Algorithm \ref{algo_bs} corresponds to the  Markov bootstrap procedure I of \citet{athreya_fuh}. For this bootstrap procedure, it is shown in Theorem 4 of the afore cited paper  that 
    \begin{equation}
\sqrt{n}\big(\widehat P^{\ast,(npara)}_{X^\ast_{t+1} \in S|X^\ast_t=x_n}  -\widehat{P}^{(npara)}_{X_{t+1} \in S|X_t=x_n} \big)   \overset{d}{\rightarrow} \mathcal{N}\Big(0, \tau^2_S\Big),
    \end{equation}
 in probability, where the variance $\tau^2_S $ can be calculated as described in Section 3 of the afore cited paper. The result follows because  
 \[ \sqrt{n}\big(\widehat{P}^{\ast,(npara)}_{S,x_n} 
 - \widehat{P}^{(npara)}_{S,x_n}\big)= \sum_{y_k \in S} \sqrt{n}\big( \widehat{P}^{(npara)}_{X^\ast_{t+1}=y_k|X^\ast_t=x_n}- \widehat{P}^{(npara)}_{X_{t+1}=y_k|X_t=x_n}\big) \]
 and the continuity of the limiting distribution.

(ii) Since $ (X_{t}(\theta),t\in \Z )$  is for every $\theta\in\Theta$ aperiodic, positive recurrent and irreducible, it has a unique stationary distribution denoted by $ \pi_i(\theta)$ for $ i\in range(X_0)$. The bootstrap time series $ X_1^\ast,\ldots, X_n^\ast$  can be considered as stemming  from the Markov chain $ (X_t(\widehat{\theta}), t \in\Z)$. Without loss of generality, assume that this bootstrap time series   starts with the stationary distribution. Then, $P(X^\ast_t=i)=\pi_i(\widehat{\theta})$.  Using  Theorem 3, Theorem 4 and Remark 4 of \citet{athreya_fuh} and in order to establish assertion (ii)  of the theorem, it suffices to show  that, as $n\rightarrow\infty$,  
\begin{enumerate}
 \item[(a)] \ $ P_{X^\ast_{t+1}=j|X_t^\ast=i} \stackrel{P}{\rightarrow} P^{(para)}_{X_{t+1}=j|X_t=i}(\theta_0)$ for every $ i,j \in range(X_0)$, and
 \item[(b)] $ P(X^\ast_{t}=i)  \stackrel{P}{\rightarrow}\pi_i (\theta_0)$ for every $ i \in range(X_0)$. 
\end{enumerate} 
Observe that condition (iii) of Theorem 3 of \citet{athreya_fuh} is satisfied since the bootstrap Markov chain $(X_t(\widehat{\theta}),t\in\Z)$ is irreducible, positive recurrent and aperiodic. Now,  (a) follows since for any $i,j\in range(X_0)$, 
\[P_{X^\ast_{t+1}=j|X_t^\ast=i}=P^{(para)}_{X_{t+1}=j|X_t=i}(\widehat{\theta}) \stackrel{P}{\rightarrow} P^{(para)}_{X_{t+1}=j|X_t=i}(\theta_0) \]
by the fact that $ \widehat{\theta}\stackrel{P}{\rightarrow} \theta_0$ and the continuity of the parametric one step transition probabilities with respect to $\theta$. Similarly,
$P(X^\ast_{t}=i)=\pi_i(\widehat{\theta}) \stackrel{P}{\rightarrow} \pi_i(\theta_0) $. \hfill $\Box$

\noindent{\bf Proof of Theorem~\ref{th.ConML}:}  
By the continuity of $ {\rm E}(\log(P^{(para)}_{X_t|X_{t-1}}(\theta))$ and in order to establish the assertion of the theorem, it suffices to show that, as $ n \rightarrow \infty$,  
\begin{equation} \label{eq.Unif-WLLN1}
\sup_{\theta\in\Theta}\left|\frac{1}{n-1}\sum_{t=2}^n \log\big(P^{(para)}_{X_t|X_{t-1}}(\theta)\big)- {\rm E} \big(\log P^{(para)}_{X_t|X_{t-1}}(\theta)\big) \right| \stackrel{P}{\rightarrow} 0,
\end{equation}
since  by Assumption \ref{assum_ML1}(i), $\theta_0$ is the unique maximum of ${\rm E} \big(\log \big( P^{(para)}_{X_t|X_{t-1}}(\theta)\big)\big) $. 
To  see why  (\ref{eq.Unif-WLLN1}) holds true,  we verify the conditions stated in  Theorem 1(a) of \citet{andrews1992}. 
Let 
\[ G_n(\theta) := \frac{1}{n-1}\sum_{t=2}^n \log \big( P^{(para)}_{X_t|X_{t-1}}(\theta)\big) -{\rm E} \big(\log \big( P^{(para)}_{X_t|X_{t-1}}(\theta)\big)\big).\]
 $ \Theta$ is  compact    by Assumption \ref{assum_p}(i).   The pointwise convergence  $ G_n(\theta) \stackrel{P}{\rightarrow} 0$ as $n\rightarrow\infty$,    for any  $\theta \in\Theta$, follows from  the ergodicity of $\{X_t,t\in\Z\}$, see Assumption \ref{assum_np}. It remains to show the stochastic equicontinuity condition, that is, Assumption SE in the afore cited theorem.  For this we have to  show  that for any $ \epsilon>0$ there exists $ \eta >0$ such that 
\begin{equation} \label{eq.EQ-Con1}
\limsup_{n\rightarrow\infty} P\Big(\sup_{\theta\in\Theta}\sup_{\widetilde{\theta} \in B(\theta,\eta)} \big|G_n(\widetilde{\theta}) -G_n(\theta)  \big|>\epsilon \Big) < \epsilon,
\end{equation}
where $ B(\theta,\eta)$ denotes the closed ball in $\Theta$ of radius $ \eta\geq 0$  centered at $ \theta$.   Notice that $ \sup_{\theta\in\Theta}  \big\|\nabla_\theta \log P^{(para)}_{X_2|X_{1}}(\theta) \|<\infty $ since $\nabla_\theta \log P^{(para)}_{X_2|X_{1}}(\theta) $ is continuous on  a compact set. By Markov's inequality and the mean value theorem, we have for some $ \theta^{\prime}$ such that $ \|\theta^{\prime}-\theta \| \leq \|\widetilde{\theta} - \theta \| $, that 
\begin{align*}
\limsup_{n\rightarrow\infty} P\Big(\sup_{\theta\in\Theta}\sup_{\widetilde{\theta} \in B(\theta,\eta)} \big|G_n(\widetilde{\theta}) -G_n(\theta)  \big|>\epsilon \Big) &
\leq \frac{1}{\epsilon} \limsup_{n\rightarrow\infty} {\rm E} \sup_{\theta\in\Theta}\sup_{\widetilde{\theta} \in B(\theta,\eta)}    \big|(\widetilde{\theta} -\theta)^\top \nabla_\theta G_n(\theta^{\prime})\big| \\
& \leq  \frac{\eta}{\epsilon} \limsup_{n\rightarrow\infty} {\rm E} \sup_{\theta\in\Theta}  \big\|\nabla_\theta G_n(\theta)\big\|\\
& 
\leq  \frac{\eta}{\epsilon} \, {\rm E}\, \sup_{\theta\in\Theta}  \big\|\nabla_\theta \log P^{(para)}_{X_2|X_{1}}(\theta) \| \\
& \ \ \ \ + \frac{\eta}{\epsilon}\sup_{\theta\in\Theta}  \big\|{\rm E}(\nabla_\theta\log P^{(para)}_{X_2|X_{1}}(\theta)) \|\\
& \leq  \frac{2\eta}{\epsilon} {\rm E}\, \sup_{\theta\in\Theta}  \big\|\nabla_\theta \log P^{(para)}_{X_2|X_{1}}(\theta) \|\\
& \leq \frac{2\eta}{\epsilon} C,
\end{align*}
for a constant $ C>0$. Thus, for $ 0< \eta < \epsilon^2/(2C)$, condition  (\ref{eq.EQ-Con1}) is satisfied. 
\hfill $\Box$

\bigskip

\noindent {\bf Proof of Theorem~\ref{th.Boot-ML-CLT}:}

(i). \ In this case, $ X_1^\ast,  \ldots, X_n^\ast$ is generated using a model from the class $ {\mathcal M}_\theta$ with estimated parameter $ \widehat{\theta}_{ML} $. 
The conditional maximum likelihood estimator $ \widehat{\theta}^\ast_{ML}$  based on this bootstrap time series  is  defined as  
\[ \widehat{\theta}^\ast_{ML} = {\rm argmax}_{\theta\in\Theta} l_n(\theta|X_1^\ast) = {\rm argmax}_{\theta\in\Theta} 
\sum_{t=2}^n \log \big( P^{(para)}_{X_t^\ast|X_{t-1}^\ast}(\theta)\big)
.\]
Note that 
\begin{equation} \label{eq.ML-Boot} \widehat{\theta}_{ML}= {\rm argmax}_{\theta\in\Theta} {\rm E}^\ast\big( 
\log \big( P^{(para)}_{X_t^\ast|X_{t-1}^\ast}(\theta)\big)={\rm argmax}_{\theta\in\Theta} {\rm E}_{\widehat{\theta}_{ML}}\big( 
\log \big( P^{(para)}_{X_t^\ast|X_{t-1}^\ast}(\theta)\big),
\end{equation}
where
\[ {\rm E}^\ast\big( 
\log \big( P^{(para)}_{X_t^\ast|X_{t-1}^\ast}(\theta)\big)=\sum_{r}\sum_{s} \log(P^{(para)}_{X_{t}^\ast=r|X^\ast_{t-1}=s}(\theta))P^{(para)}_{\widehat{\theta}_{ML}}(X_t^\ast=r,X^\ast_{t-1}=s) \]
and the double summation extends over the entire range of $X_0$.
We use the expansion
\begin{equation}  \label{eq.A6} \frac{1}{n-1}l^{''}_n(\widetilde{\theta}^\ast_n|X_1^\ast)\sqrt{n-1}\big(\widehat{\theta}^\ast_{ML}-\widehat\theta_{ML}\big) = -\frac{1}{\sqrt{n-1}}l^{'}_n(\widehat{\theta}_{ML}|X_1^\ast),
\end{equation}
see also (\ref{eq.ML-exp1}), 
for some $\widetilde{\theta}^\ast_n $ such that  $ \|\widetilde{\theta}^\ast_n-\widehat\theta_{ML}\| \leq \|\widehat{\theta}^\ast_{ML}-\widehat\theta_{ML}\|$. Let
\begin{align*}
    \nabla_\theta \log(P^{(para)}_{X^\ast_t|X^\ast_{t-1}}(\theta) \big)\big|_{\theta=\theta^\prime}& =: g(X^\ast_t,X_{t-1}^\ast; \theta^\prime ).
\end{align*}
Then, we have
\begin{equation}
    \label{eq.MDArray-1} \frac{1}{\sqrt{n-1}}l^{'}_n(\widehat{\theta}_{ML}|X_1^\ast) = \frac{1}{\sqrt{n-1}} \sum_{t=2}^{n} g(X_t^\ast,X_{t-1}^\ast; \widehat{\theta}_{ML}),
\end{equation}
and (\ref{eq.MDArray-1}) is (conditional on $X_1,\ldots, X_n$) a sequence of martingale differences. This holds true  since 
\begin{align} \label{eq.MD-Array}
    {\rm E}^\ast\big( \nabla_\theta  & \log(P^{(para)}_{X^\ast_t|X^\ast_{t-1}}(\theta) \big)\big|_{\theta=\widehat{\theta}_{ML} }\big| X_{t-1}^\ast\big)  \\
    & 
    =\sum_r \frac{1}{P^{(para)}_{X_t^\ast=r|X_{t-1}^\ast}(\widehat{\theta}_{ML})}\nabla_\theta P^{(para)}_{X_t^\ast=r|X_{t-1}^\ast}(\theta)\big|_{\widehat{\theta}_{ML}} P^{(para)}_{X_t^\ast=r|X_{t-1}^\ast}(\widehat{\theta}_{ML})  \nonumber \\
    & =\sum_r \nabla_\theta P^{(para)}_{X_t^\ast=r|X_{t-1}^\ast}(\widehat{\theta}_{ML})   \nonumber \\
    & = \nabla_\theta \Big(\sum_r P^{(para)}_{X_t^\ast=r|X_{t-1}^\ast}(\widehat{\theta}_{ML})\Big) \nonumber  \\
    & =0, \nonumber
\end{align}
where all  sums go  over the range of $X_t^\ast$.
Notice that by (\ref{eq.ML-Boot}) and the exchangeability of differentiation and expectation, see Assumption 4(iii), it is easily seen that  $ {\rm E}^\ast(g(X_t^\ast,X_{t-1}^\ast; \widehat{\theta}_{ML}))=0$. 

For the matrix of second-order partial derivatives and  since the elements of the 
matrix  $ \nabla^2_\theta \log(P^{(para)}_{X_t|X_{t-1}}(\theta)\big)$ are Lipschitz continuous functions of $ \theta$, we get  for the Frobenius norm $ \|\cdot\|_F$, 
\begin{align} \label{eq.Theta-dif}
\Big\| \Big(\frac{1}{n-1}l^{''}_n(\widetilde{\theta}^\ast_n|X_1^\ast)\Big)-\Big(\frac{1}{n-1}l^{''}_n(\theta_0|X_1^\ast)\Big)  \Big\|_F & \leq \Big\| \Big(\frac{1}{n-1}l^{''}_n(\widetilde{\theta}^\ast_n|X_1^\ast)\Big)-\Big(\frac{1}{n-1}l^{''}_n(\widehat\theta_{ML}|X_1^\ast)\Big)  \Big\|_F \nonumber \\
& \ \ \ \ + \Big\| \Big(\frac{1}{n-1}l^{''}_n(\widehat\theta_{ML}|X_1^\ast)\Big)-\Big(\frac{1}{n-1}l^{''}_n(\theta_0|X_1^\ast)\Big)  \Big\|_F \nonumber \\
& \leq  O_{P}\big(\|\widehat{\theta}^\ast_{ML}-\widehat\theta_{ML}\| \big) + O_{P}\big(\|\widehat\theta_{ML}-\theta_{0}\| \big).
\end{align}
The second term of the last inequality goes to zero because 
$\widehat\theta_{ML} \stackrel{P}{\rightarrow} \theta_{0} $.
The first  term of  (\ref{eq.Theta-dif})  vanishes asymptotically  by the triangular inequality and  because $ \|\widehat{\theta}_{ML}-\theta_{0} \| \stackrel{P}{\rightarrow} 0$ and  $ \|\widehat{\theta}^\ast_{ML}-\theta_{0} \| \stackrel{P}{\rightarrow} 0$, in probability. To establish the latter statement, it suffices to   show that 
\begin{equation} \label{eq.ULLN-2}
\sup_{\theta \in \Theta}\Big|\frac{1}{n-1}\sum_{t=2}^n \log  P^{(para)}_{X_t^\ast|X_{t-1}^\ast}(\theta)- {\rm E}_{\theta_0} \big(\log \big( P^{(para)}_{X_t|X_{t-1}}(\theta)\big)\big) \Big| \stackrel{P}{\rightarrow} 0,\end{equation}
where
\[{\rm E}_{\theta_0}\big( 
\log \big( P^{(para)}_{X_t|X_{t-1}}(\theta)\big)=\sum_{r}\sum_{s} \log P^{(para)}_{X_{t}=r|X_{t-1}=s}(\theta)P^{(para)}_{{\theta}_{0}}(X_t=r,X_{t-1}=s). \]
Notice that in the above set up, $ (X_t,X_{t-1})$  follows the distribution  generated by a parametric  model with parameter $ \theta_0$ while $ (X_t^\ast, X_{t-1}^\ast)$ that of the same model but with parameter $ \widehat{\theta}_{ML}$.

Using the notation 
\[ W^\ast_n(\theta):=\frac{1}{n-1}\sum_{t=2}^n \log P^{(para)}_{X_t^\ast|X_{t-1}^\ast}(\theta)- {\rm E}_{\theta_0} \big(\log \big( P^{(para)}_{X_t|X_{t-1}}(\theta)\big)\big) \]
and arguing as in the proof of (\ref{eq.Unif-WLLN1}), it suffices to show that  $ W_n^\ast (\theta) \rightarrow 0$ in probability for any $\theta \in \Theta$  and that $ \eta>0$ exists such that, in probability,  
\begin{equation} \label{eq.EQ-Con2}
\limsup_{n\rightarrow\infty} P\Big(\sup_{\theta\in\Theta}\sup_{\widetilde{\theta} \in B(\theta,\eta)} \big|W^\ast_n(\widetilde{\theta}) -W^\ast_n(\theta)  \big|>\epsilon \Big) < \epsilon,
\end{equation}
for all $ \epsilon>0$. We have
\begin{align*}
\limsup_{n\rightarrow\infty} {\rm E}^\ast \sup_{\theta\in\Theta} & \sup_{\widetilde{\theta}\in B(\theta,\eta)}\big| W_n^\ast(\widetilde{\theta}) - W_n^\ast(\theta)\big| \\
& \leq \eta\limsup_{n\rightarrow\infty} {\rm E}^\ast \sup_{\theta\in\Theta}\|(n-1)^{-1}\sum_{t=2}^n \nabla_\theta \log P_{X^\ast_t|X^\ast_{t-1}}\|\\
& \ \ \ \ + \eta \sup_{\theta\in\Theta} \|{\rm E}_{\theta_0} (\nabla_\theta \log P^{(para)}_{X_2|X_1}(\theta))\|\\
& \leq \eta \limsup_{n\rightarrow\infty} \sum_r\sum_s \sup_{\theta\in\Theta} \|\nabla_\theta \log P^{(para)}_{X_2=r|X_1=s}(\theta)\| P^{(para)}_{\widehat{\theta}_{ML}}(X_2=r,X_1=s) \\
& \ \ \ \ + \eta \sup_{\theta\in\Theta} \|{\rm E}_{\theta_0} (\nabla_\theta \log P^{(para)}_{X_2|X_1}(\theta))\|\\
&=\eta  \sum_r\sum_s \sup_{\theta\in\Theta} \|\nabla_\theta \log P^{(para)}_{X_2=r|X_1=s}(\theta)\| P^{(para)}_{\theta_{0}}(X_2=r,X_1=s)\\
& \ \ \ \ + \eta \sup_{\theta\in\Theta} \|{\rm E}_{\theta_0} \nabla_\theta \log P^{(para)}_{X_2|X_1}(\theta)\| \\
& \leq 2 \eta\,  {\rm E}_{\theta_0}\big(\sup_{\theta\in\Theta}\| \nabla_\theta \log P^{(para)}_{X_2|X_1}(\theta)\|\big).
\end{align*}
Note that the equality before the last inequality follows because $ \widehat{\theta}_{ML} \stackrel{P}{\rightarrow} \theta_0$  implies $ P^{(para)}_{\widehat{\theta}_{ML}}(X_2=r,X_1=s)  \stackrel{P}{\rightarrow} P^{(para)}_{{\theta}_{0}}(X_2=r,X_1=s)$  for every $r,s$. This together with the boundedness in probability   of $\sup_{\theta\in\Theta} \|\nabla_\theta \log P^{(para)}_{X_2=r|X_1=s}(\theta)\|$ 
leads to 
\begin{align*} 
\sum_r\sum_s \sup_{\theta\in\Theta} \|\nabla_\theta & \log P^{(para)}_{X_2=r|X_1=s}(\theta)\|\big| P^{(para)}_{\widehat{\theta}_{ML}}(X_2=r,X_1=s)- P^{(para)}_{\theta_{0}}(X_2=r,X_1=s)\big|\\
& \leq O_P(1) \sum_r\sum_s \big|P^{(para)}_{\widehat{\theta}_{ML}}(X_2=r,X_1=s)-P^{(para)}_{\theta_{0}}(X_2=r,X_1=s)\big|\stackrel{P}{\rightarrow} 0, 
\end{align*}
where the last convergence follows by  Scheffé's Theorem. Therefore, for  $ 0<\eta<\epsilon^2/2C$, where $ C={\rm E}_{\theta_0}\big(\sup_{\theta\in\Theta}\| \nabla_\theta \log P^{(para)}_{X_2|X_1}(\theta)\|\big)<\infty$,   assertion (\ref{eq.EQ-Con2}) follows.

By the weak  law of large numbers and (\ref{eq.Theta-dif}),  we have  
\begin{align*}
    \Big\| {\rm E}\big(\nabla^2_\theta \log(P^{(para)}_{X_t|X_{t-1}}(\theta)\big)\big|_{\theta=\theta_0}\big) & -\Big(\frac{1}{n-1}l^{''}_n(\widetilde{\theta}^\ast_n|X_1^\ast)\Big)\Big\|_F    \\
    & \leq   \Big\| E\big(\nabla^2_\theta \log(P^{(para)}_{X_t|X_{t-1}}(\theta)\big)\big|_{\theta=\theta_0}\big) -\Big(\frac{1}{n-1}l^{''}_n(\theta_0|X_1)\Big) \Big\|_F \\
    & \ \ \ \ + \Big\| \Big(\frac{1}{n-1}l^{''}_n(\theta_0|X_1)\Big) -\Big(\frac{1}{n-1}l^{''}_n(\widehat\theta^\ast_{ML} |X^\ast_1)\Big)  \Big\|_F\\
    & \ \ \ \ 
    + \Big\| \Big(\frac{1}{n-1}l^{''}_n(\widehat\theta_{ML}^\ast|X^\ast_1)\Big) -\Big(\frac{1}{n-1}l^{''}_n(\widetilde{\theta}_n^\ast |X^\ast_1)\Big)  \Big\|_F\\ 
    & \stackrel{P}{\rightarrow} 0. 
\end{align*}
Invoking a central limit theorem for triangular arrays of martingale differences, see  \citet{brown1971} and \citet{alj},  to the sequence 
\[\frac{1}{\sqrt{n-1}} \sum_{t=2}^{n} \big(g(X_t^\ast,X_{t-1}^\ast; \widehat{\theta}_{ML}\big) - E^\ast\big( g(X_t^\ast,X_{t-1}^\ast, ;\hat{\theta}_{ML}\big)\big),\] 
also see (\ref{eq.MDArray-1}), and using the fact that the matrix 
\[  {\rm E}_{\theta_0}\big(\nabla^2_\theta \log(P^{(para)}_{X_t|X_{t-1}}(\theta)\big)\big|_{\theta=\theta_0}\big),\]
is positive definite, see Assumption \ref{assum_p}, the assertion of the theorem follows. 

\bigskip

Proof of assertion  (ii).  The proof of this assertion essentially follows the arguments used for establishing assertion (i) of the theorem with the main  difference concerning  the CLT used.
Similar to (\ref{eq.A6}), we have 
\begin{equation}\label{eq.MDArray-1_ii}  \frac{1}{n-1}l^{''}_n(\widetilde{\theta}^\ast_n|X_1^\ast)\sqrt{n-1}\big(\widehat{\theta}^\ast_{ML}-\widehat\theta_{ML}\big) = -\frac{1}{\sqrt{n-1}}l^{'}_n(\widehat{\theta}_{ML}|X_1^\ast),
\end{equation}
for some $\widetilde{\theta}^\ast_n $ such that  $ \|\widetilde{\theta}^\ast_n-\widehat\theta_{ML}\| \leq \|\widehat{\theta}^\ast_{ML}-\widehat\theta_{ML}\|$.
Notice  that  the centering here  by $ \widehat{\theta}_{ML} $ is justified by the fact that,
\begin{align*} 
 {\rm argmax}_{\theta\in\Theta} {\rm E}^\ast(\log P_{X^\ast_t|X^\ast_{t-1}}(\theta)) 
&= {\rm argmax}_{\theta\in\Theta} \sum_{r}\sum_{s}\log P_{X_{t}=r|X_{t-1}=s}(\theta) \widehat{P}^{(npara)}_{X_{t}=r|X_{t-1}=s}\\
&= {\rm argmax}_{\theta\in\Theta} \sum_{t=2}^{n} \log P_{X_t|X_{t-1}}(\theta) =\widehat{\theta}_{ML}.
\end{align*}
According to \eqref{eq.MDArray-1}, in \eqref{eq.MDArray-1_ii} as well, we have
\begin{equation}\label{eq.represenatation_Y_tn_star}
\frac{1}{\sqrt{n-1}}l^{'}_n(\widehat{\theta}_{ML}|X_1^\ast) = \frac{1}{\sqrt{n-1}} \sum_{t=2}^{n} g(X_t^\ast,X_{t-1}^\ast; \widehat{\theta}_{ML}) =: \frac{1}{\sqrt{n-1}} \sum_{t=2}^{n} Y_{t,n}^\ast,
\end{equation}
where the pseudo time series $ X_1^\ast, \ldots, X_n^\ast$ is generated using the non-parametrically estimated one step transition probabilities $\widehat P^{(npara)}_{X_{t+1}=x_j|X_t=x_i}$  for $ x_i,x_j\in range(X_0)$ (instead of the parametrically estimated one step transition probabilities $P^{(para)}_{X_{t+1}=x_j|X_t=x_i}(\widehat\theta_{\text{ML}})$ used in part (i)). In the following, we will use the central limit theorem of \citet[Theorem 1]{peligrad2012} for triangular arrays of non-homogeneous Markov chains and adapt it to our (bootstrap) setup. Note that here we are dealing with the homogeneous case only, which simplifies  arguments. 
By forming the two-dimensional Markov chain $((X_t,X_{t-1})^\prime,t\in\mathbb{Z})$, the latter shares important properties with $(X_t,t\in\mathbb{Z})$. In particular, the corresponding maximal correlation coefficient is still strictly smaller than $1$. 

Moreover, conditional on $X_1,\ldots,X_n$, the bootstrap process $(X_t^\ast,t\in\mathbb{Z})$
 fulfills these properties as well with probability tending to one. We have \begin{align*}
    \rho_1^* = \sup_{f,g}\left\{\frac{\big|E^*(f(X_i^*)g(X_{i-1}^*))-E^*(f(X_i^*))E^*(g(X_{i-1}^*))\big|}{\sqrt{E^*(f^2(X_i^*))}\sqrt{E^*(g^2(X_{i-1}^*))}}; \|f(X_i^*)\|_2^*<\infty,\; \|g(X_{i-1}^*)\|_2^*<\infty\right\},
\end{align*}
where we used the notation $\|X^*\|_p^*=(E^*((X^*)^p))^{1/p}$ for $p>1$. We show that, as $n\rightarrow \infty$, $\rho_1^*\rightarrow  \rho_1$ in probability holds true, where $\rho_1$ is the corresponding coefficient of the underlying Markov process which  is assumed to be strictly smaller than 1. After having achieved this, we can argue that $\rho_1^*<1$ holds true in probability and we can make use of \citet[Theorem 1]{peligrad2012} to establish a bootstrap CLT for \eqref{eq.represenatation_Y_tn_star}. Toward this goal, we show that
\begin{align}
    |\rho_1^*-\rho_1|\rightarrow 0, 
\end{align}
in probability, as $n\rightarrow \infty$. Consider the nominators of $\rho_1^*$ and $\rho_1$. Then, we have
\begin{align*}
    & |(E^*(f(X_i^*)g(X_{i-1}^*))-E^*(f(X_i^*))E^*(g(X_{i-1}^*)))-(E(f(X_i)g(X_{i-1}))-E(f(X_i))E(g(X_{i-1})))| \\
    \leq & |E^*(f(X_i^*)g(X_{i-1}^*))-E(f(X_i)g(X_{i-1}))| + |E^*(f(X_i^*))E^*(g(X_{i-1}^*))-E(f(X_i))E(g(X_{i-1}))|.
\end{align*}
We show that $|E^*(f(X_i^*)g(X_{i-1}^*))-E(f(X_i)g(X_{i-1}))| \rightarrow 0$ and $|E^*(f(X_i^*))E^*(g(X_{i-1}^*))-E(f(X_i))E(g(X_{i-1}))|\rightarrow 0$, where the latter is implied by $|E^*(f(X_i^*))-E(f(X_i))|\rightarrow 0$. Similarly, we can deal with the denominator, where we have to show that $|E^*(f^2(X_i^*))-E(f^2(X_i))|\rightarrow 0$. We only elaborate on  the most cumbersome term $|E^*(f(X_i^*)g(X_{i-1}^*))-E(f(X_i)g(X_{i-1}))|$ (all other terms can be dealt with analogously). We  show that
\begin{align*}
E^*(f(X_i^*)g(X_{i-1}^*)) \rightarrow E(f(X_i)g(X_{i-1}))    
\end{align*}
in probability,  as $n\rightarrow \infty$. For this purpose, let ${\mathcal N}$ denote the state space of $(X_t,t\in\mathbb{Z})$ and ${\mathcal N}^\ast_n$ the state space of $(X_t^*,t\in\mathbb{Z})$. Note that, for all $n\in\mathbb{N}$, we have ${\mathcal N}_n^* \subseteq {\mathcal N}$ with ${\mathcal N}_n^*$ being finite. 
Then, 
\begin{align*}
E^*(f(X_i^*)g(X_{i-1}^*)) = & \sum_{r,s\in {\mathcal N}_n^*}f(r)g(s)P^*(X_i^*=r,X_{i-1}^*=s) \\
= & \sum_{r,s\in {\mathcal N}_n^*}f(r)g(s)\left(P^*(X_i^*=r,X_{i-1}^*=s)-P(X_i=r,X_{i-1}=s)\right) \\
&+ \sum_{r,s\in {\mathcal N}_n^*}f(r)g(s)P(X_i=r,X_{i-1}=s)    \\
=& I_n+II_n.
\end{align*}
As ${\mathcal N}_n^*\rightarrow {\mathcal N}$ monotonically as $n\rightarrow \infty$, we get by the  monotone convergence theorem  that $II_n\rightarrow E(f(X_i)g(X_{i-1}))$ in probability. It remains to show that $I_n=o_p(1)$. We have
\begin{align*}
I_n
=& \sum_{r,s\in {\mathcal N}_n^*}f(r)g(s)\left(P^*(X_i^*=r,X_{i-1}^*=s)-P(X_i=r,X_{i-1}=s)\right)  \\
=& \sum_{r,s\in {\mathcal N}}f(r)g(s)\left(P^*(X_i^*=r,X_{i-1}^*=s)-P(X_i=r,X_{i-1}=s)\right)\mathbb{1}(P^*(X^*=r)P^*(X^*=s)>0)  \\
=& \sum_{r,s\in {\mathcal N}}\bigg[f(r)g(s)\mathbb{1}(P^*(X^*=r)P^*(X^*=s)>0)\bigg]\left(P^*(X_i^*=r,X_{i-1}^*=s)-P(X_i=r,X_{i-1}=s)\right)    \\
\leq& \sum_{r,s\in {\mathcal N}}\bigg|f(r)g(s)\mathbb{1}(P^*(X^*=r)P^*(X^*=s)>0)\bigg|\cdot\left|P^*(X_i^*=r,X_{i-1}^*=s)-P(X_i=r,X_{i-1}=s)\right|.
\end{align*}
Furthermore,
\begin{align*}
& \bigg|f(r)g(s)\mathbb{1}(P^*(X^*=r)P^*(X^*=s)>0)\bigg|    \\
\leq & \sup_{r,s}\bigg|f(r)g(s)\mathbb{1}(P^*(X^*=r)P^*(X^*=s)>0)\bigg|    \\
\leq & \sup_{r}\bigg|f(r)\mathbb{1}(P^*(X^*=r)>0)\bigg| \sup_{s}\bigg|g(s)\mathbb{1}(P^*(X^*=s)>0)\bigg|    \\
\leq & \left(\max\{|f(X_1)|,\ldots,|f(X_n)|\}\right)^2.
\end{align*}
Hence, it suffices to show that
\begin{align*}
\left(\max\{|f(X_1)|,\ldots,|f(X_n)|\}\right)^2\sum_{r,s\in {\mathcal N}}\left|P^*(X_i^*=r,X_{i-1}^*=s)-P(X_i=r,X_{i-1}=s)\right|=o_P(1).
\end{align*}
For this purpose, we aim to show
\begin{align}
\frac{\left(\max\{|f(X_1)|,\ldots,|f(X_n)|\}\right)^2}{\sqrt{n}} =& o_P(1)  \label{factor1} \\
\text{and} \,\sqrt{n}\sum_{r,s\in {\mathcal N}}\left|P^*(X_i^*=r,X_{i-1}^*=s)-P(X_i=r,X_{i-1}=s)\right| =& O_P(1)  \label{factor2}
\end{align}
in the following. Starting with term \eqref{factor1}, for any $x>0$, we have
\begin{align*}
P\left(\frac{\left(\max\{|f(X_1)|,\ldots,|f(X_n)|\}\right)^2}{\sqrt{n}}>x\right) =& P\left(\max\{|f(X_1)|,\ldots,|f(X_n)|\}>x^{1/2}n^{1/4}\right)   \\
=& P\left(\bigcup_{i=1}^n\left\{|f(X_i)|>x^{1/2}n^{1/4}\right\}\right)    \\
\leq& \sum_{i=1}^n P\left(\left\{|f(X_i)|>x^{1/2}n^{1/4}\right\}\right)    \\
=& n P\left(\left\{|f(X_1)|^k>(x^{1/2}n^{1/4})^k\right\}\right)    \\
\leq& n \frac{E(|f(X_1)|^k)}{(x^{1/2}n^{1/4})^k}    \\
=& \frac{n^{1-k/4}}{x^{k/2}}E(|f(X_1)|^k).
\end{align*}
Hence, for $k>4$, the last right-hand side converges to zero in probability, if ${\rm E}(|f(X_1)|^k)<\infty$ holds for all square-integrable functions $f$. That is, we need $8+\delta$ moments, which holds as we have assumed $12+\delta$ moments for the process $(X_t,t\in\mathbb{Z})$.

Continuing with \eqref{factor2}, using $S=\{x_i,i\in\mathbb{N}\}$, we have
\begin{align*}
& \sqrt{n}\sum_{r,s\in {\mathcal N}} |\widehat P(X_1=r,X_0=s)-P(X_1=r,X_0=s)|  \\
\leq & \sqrt{n}\sum_{i,j=1}^K |\widehat P(X_1=x_i,X_0=x_j)-P(X_1=x_i,X_0=x_j)|   \\
& + \sqrt{n}\sum_{i=K+1}^\infty\sum_{j=1}^\infty |\widehat P(X_1=x_i,X_0=x_j)-P(X_1=x_i,X_0=x_j)|   \\
& + \sqrt{n}\sum_{i=1}^\infty\sum_{j=K+1}^\infty |\widehat P(X_1=x_i,X_0=x_j)-P(X_1=x_i,X_0=x_j)|   \\
= & I_{K,n} + II_{K,n} + III_{K,n},
 \end{align*}
where $K$ is some arbitrarily large positive integer. For all fixed $K$, we have that  $I_{K,n}=O_P(1)$, because a multivariate CLT holds for all finitely many relative frequencies, when properly centered and scaled with $\sqrt{n}$. Further, the second term $II_{K,n}$ can be bounded by
\begin{align*}
    & \sqrt{n}\sum_{i=K+1}^\infty\sum_{j=1}^\infty |\widehat P(X_1=x_i,X_0=x_j)| + \sqrt{n}\sum_{i=K+1}^\infty\sum_{j=1}^\infty |P(X_1=x_i,X_0=x_j)|  \\
    \leq & \sqrt{n}\sum_{i=K+1}^\infty\widehat P(X_1=x_i) + \sqrt{n}\sum_{i=K+1}^\infty P(X_1=x_i)  \\
    = & \sqrt{n}(1- \sum_{i=1}^K \widehat P(X_1=x_i)) + \sqrt{n}\sum_{i=K+1}^\infty P(X_1=x_i)    \\
    = & \sqrt{n}(1- \sum_{i=1}^K \big(\widehat P(X_1=x_i) - P(X_1=x_i)\big)) -\sqrt{n}\sum_{i=1}^K P(X_1=x_i) + \sqrt{n}\sum_{i=K+1}^\infty P(X_1=x_i)    \\
    \leq & \sqrt{n}(1- \sum_{i=1}^K \big(\widehat P(X_1=x_i) - P(X_1=x_i)\big)) -\sqrt{n}\bigg(1-\sum_{i=K+1}^\infty P(X_1=x_i)\bigg) + \sqrt{n}\sum_{i=K+1}^\infty P(X_1=x_i)   \\
    = & -\sqrt{n}\sum_{i=1}^K \big(\widehat P(X_1=x_i) - P(X_1=x_i)\big) +\sqrt{n}\sum_{i=K+1}^\infty P(X_1=x_i) + \sqrt{n}\sum_{i=K+1}^\infty P(X_1=x_i)   \\
    \leq & O_P(1) +2\sqrt{n}\sum_{i=K+1}^\infty P(X_1=x_i)
\end{align*}
for all $K$ sufficiently large. Analogously, we get the same for term $III_{K,n}$. Hence, altogether, we have
\begin{align*}
\frac{\left(\max\{|f(X_1)|,\ldots,|f(X_n)|\}\right)^2}{\sqrt{n}} O_P(1) =o_P(1)   
\end{align*}
and it remains to argue that, for all $K$, we also have
\begin{align*}
\frac{\left(\max\{|f(X_1)|,\ldots,|f(X_n)|\}\right)^2}{\sqrt{n}}2\sqrt{n}\sum_{i=K+1}^\infty P(X_1=x_i)=o_P(1).   
\end{align*}
By the same arguments used above for \eqref{factor1}, we get
\begin{align*}
& \frac{\left(\max\{|f(X_1)|,\ldots,|f(X_n)|\}\right)^2}{\sqrt{n}}2\sqrt{n}\sum_{i=K+1}^\infty P(X_1=x_i)   \\
=& O_P\left(n^{1-k/4}{\rm E}(|f(X_1)|^k)\sqrt{n}\sum_{i=K+1}^\infty P(X_1=x_i)\right) \\
=& O_P\left(n^{(6-k)/4}{\rm E}(|f(X_1)|^k)\sum_{i=K+1}^\infty P(X_1=x_i)\right).
\end{align*}
Hence, for $k>6$, the last right-hand side converges to zero in probability, if ${\rm E}(|f(X_1)|^k)<\infty$ holds for all square-integrable functions $f$. This is the case, as we have assumed $12+\delta$ moments for the process $(X_t,t\in\mathbb{Z})$.

Now since $\rho_1^\ast<1$ and having established that $\rho_1^* \rightarrow \rho_1$, in probability,   there exists a $\widetilde\rho_1$ with $\rho_1<\widetilde\rho_1<1$ such that, with high probability, $\rho_1^*<\widetilde \rho_1<1$ holds true. Hence, we can  use  Theorem 1 in \citet{peligrad2012}. According to \eqref{eq.represenatation_Y_tn_star}, we are concerned with
\begin{align}
    L_n^* := \sum_{t=2}^n Y_{t,n}^*,
\end{align}
where
\begin{align}
    Y_{t,n}^*  = g_n(X_t^\ast,X_{t-1}^\ast) := g(X_t^\ast,X_{t-1}^\ast; \widehat{\theta}_{ML}) = \nabla_\theta \log(P^{(para)}_{X^\ast_t|X^\ast_{t-1}}(\theta) \big)\big|_{\theta=\widehat \theta_{ML}}.
\end{align}
Moreover, as shown above, we have ${\rm E}^*(Y_{t,n}^*)=0$ as well as ${\rm E}^*((Y_{t,n}^*)^2)<\infty$ and let  $\widehat\sigma_n^{2}:={\rm Var}^*(L_n^*)$. Then, conditional on $X_1,\ldots,X_n$, we have
\begin{align}
    \max_{1\leq t\leq n} |Y_{t,n}^*| = \max_{1\leq t\leq n} \nabla_\theta \log(P^{(para)}_{X^\ast_t|X^\ast_{t-1}}(\theta) \big)\big|_{\theta=\widehat \theta_{ML}} =: C_n^*(\widehat \theta_{ML}),
\end{align}
where, for each fixed $n$ and conditionally  on $\widehat \theta_{ML}$, $C_n^*=C_n^*(\widehat \theta_{ML})$ is actually a finite (non-random) constant as $X_t^*\leq \max\{X_1,\ldots,X_n\}$ by construction of the non-parametric bootstrap procedure. Also by adapting the notation in \citet{peligrad2012}, let
\begin{align}
    \widehat \lambda_n := 1-\widehat \rho_{n,1} := 1-\rho_{1}^*.
\end{align}
Then, it remains to show that
\begin{align}
    \frac{C_n^*(1+|\log(\widehat \lambda_n)|)}{\widehat \lambda_n\widehat \sigma_n} \overset{P}{\rightarrow} 0   \label{eq:condition_Peligrad}
\end{align}
as $n\rightarrow\infty$, in order  to prove that
\begin{align*}
    \frac{L_n^*}{\widehat \sigma_n} \overset{d}{\rightarrow} \mathcal{N}(0,1)   \quad   \text{in probability.}
\end{align*}
Making use of the arguments from above, condition \eqref{eq:condition_Peligrad} holds. To see this, note that $\widehat \lambda_n>0$ in probability as $\widehat \lambda_n<1$ in probability, where the latter holds, because the bootstrap process $(X_t^*,t\in\mathbb{Z})$ shares the property of a maximal correlation coefficient smaller than 1 (in probability) with the process $(X_t,t\in\mathbb{Z})$, who fulfills this by assumption. Hence, we have
\begin{align}
     \frac{C_n^*(1+|\log(\widehat \lambda_n)|)}{\widehat \lambda_n\widehat \sigma_n} = O_P\left(\frac{C_n^*}{\widehat \sigma_n}\right)
     = o_P(1)
\end{align}
by assumption \eqref{assumption_theorem_3.2(ii)} as $1/\widehat \sigma_n=O_p(1/\sqrt{n})$.
Finally, we get
\begin{align*}
    -\frac{1}{\sqrt{n-1}}l^{'}_n(\widehat{\theta}_{ML}|X_1^\ast) = - \left(\frac{1}{n-1}\widehat \sigma_n^2\right)^{1/2}\frac{L_n^*}{\widehat \sigma_n} \ \overset{d}{\rightarrow} \ \mathcal{N}(0, {\rm E}\big(\nabla_\theta^2 \log(P^{(para)}_{X_t|X_{t-1}}(\theta)\big)\big|_{\theta=\theta_0}\big)), 
\end{align*}
because of $\frac{1}{n-1}\widehat \sigma_n^2 \overset{P}{\rightarrow} {\rm E}\big(\nabla_\theta^2 \log(P^{(para)}_{X_t|X_{t-1}}(\theta)\big)\big|_{\theta=\theta_0}\big)$ by the weak law of large numbers. Finally, similar to the arguments employed for part (i), we also have 
\begin{align*}
    \frac{1}{n-1}l^{''}_n(\widetilde{\theta}^\ast_n|X_1^\ast) \overset{P}{\rightarrow} {\rm E}\big(\nabla_\theta^2 \log(P^{(para)}_{X_t|X_{t-1}}(\theta)\big)\big|_{\theta=\theta_0}\big),
\end{align*}
which completes the proof of part (ii) as ${\rm E}\big(\nabla_\theta^2 \log(P^{(para)}_{X_t|X_{t-1}}(\theta)\big)\big|_{\theta=\theta_0}\big)$ is positive definite by Assumption \ref{assum_ML1}.

\hfill $\Box$

\end{document}